\documentclass[11pt]{article}
\usepackage[utf8]{inputenc}
\usepackage[T1]{fontenc}
\usepackage{jheppub}
\usepackage{graphicx}
\usepackage{grffile}
\usepackage{longtable}
\usepackage{wrapfig}
\usepackage{rotating}
\usepackage[normalem]{ulem}
\usepackage{amsmath}
\usepackage{textcomp}
\usepackage{amssymb}
\usepackage{capt-of}
\usepackage{hyperref}
\usepackage{color}

\usepackage{subcaption}

\usepackage{tikz}
\usepackage{slashed}
\usetikzlibrary{math}
\usetikzlibrary{positioning}% To get more advances positioning options
\usetikzlibrary{arrows}% To get more arrow heads
\newcommand{\dd}{\mathrm{d}}

\author[1,2]{Samuel Abreu,}
\author[1]{Pier Francesco Monni,}
\author[3]{Ben Page,}
\author[1,4]{Johann Usovitsch}

\affiliation[1]{Theoretical Physics Department, CERN, 1211 Geneva,
  Switzerland}
\affiliation[2]{Higgs Centre for Theoretical Physics, School of
  Physics and Astronomy,\\The University of Edinburgh, Edinburh EH9
  3FD, Scotland, UK}
\affiliation[3]{Department of Physics and Astronomy, Ghent University,
  9000 Ghent, Belgium}
\affiliation[4]{Institut f\"ur Physik und IRIS Adlershof,
  Humboldt-Universit\"at zu Berlin, 10099 Berlin, Germany}

\preprint{\begin{flushright}
    CERN-TH-2024-221, HU-EP-24/40-RTG
\end{flushright}}

\date{\today}
\title{Planar Six-Point Feynman Integrals for
  Four-Dimensional Gauge Theories}

\abstract{We compute all planar two-loop six-point Feynman integrals
  entering scattering observables in massless gauge theories such as
  QCD.
  A central result of this paper is the formulation of the
  differential-equations method under the algebraic constraints
  stemming from four-dimensional kinematics, which in this case leaves
  only 8 independent scales.
  We show that these constraints imply that one must compute
  topologies with only up to 8 propagators, instead of the expected
  9. This leads to the decoupling of entire classes of integrals that
  do not contribute to scattering amplitudes in four dimensional gauge
  theories.
  We construct a pure basis and derive their canonical differential
  equations, of which we discuss the numerical solution.
  This work marks an important step towards the calculation of massless
  $2\to 4$ scattering processes at two loops.  }

\begin{document}

\maketitle

\section{Introduction}

With the upcoming high-luminosity phase of the LHC physics program only a few
years away, demands for precise theoretical predictions naturally increase. 
In recent years,
with improvements in both our ability to compute scattering amplitudes as well
as the corresponding real-radiation contributions, computations at 
next-to-next-to-leading order (NNLO)
precision are becoming commonplace.
On the side of the scattering amplitude, one major bottleneck to the computation
of new two-loop results is the availability of the associated
Feynman integrals.
Currently, the multiplicity frontier is firmly established at five points, with
the complete computations of five-point massless~\cite{Papadopoulos:2015jft,
Gehrmann:2015bfy,Gehrmann:2018yef,Abreu:2018rcw,Chicherin:2018mue,Abreu:2018aqd,Chicherin:2018old,Chicherin:2020oor} and
one-mass
integrals~\cite{Papadopoulos:2015jft,Abreu:2020jxa,Abreu:2021smk,Abreu:2023rco,Kardos:2022tpo,Canko:2020ylt,Chicherin:2021dyp},
as well as first results for two-mass planar integrals~\cite{Abreu:2024yit}.
Some new results have also recently appeared for topologies with internal 
masses, such as planar integrals for 
$t\overline{t}j$-production~\cite{Badger:2022hno,Badger:2024fgb} as well as 
a number of results for integrals for $t\overline{t}H$
production~\cite{FebresCordero:2023pww,Agarwal:2024jyq}.
Beyond five-points, 
initial results for two-loop planar six-point Feynman integrals are
also available, such as the numerical evaluation of the ``hexa-box''
family~\cite{Liu:2021wks}, the differential
equation for all planar integrals on the 
maximal-cut~\cite{Henn:2021cyv}, as well as the complete calculation of 
a subset of the integral topologies~\cite{Henn:2024ngj}.
Indeed, the computation of two-loop six-point Feynman integrals is a
necessary ingredient for a number of investigations of
phenomenological relevance for hadron colliders, 
such as obtaining more precise predictions for the production 
of four jets or for final states involving jets and photons. 
Importantly, six-point Feynman integrals
present new theoretical properties, and in order to be able to tackle
such complicated processes these must be thoroughly understood.

The first property is related to the number of external momenta.
At six points, we have for the first time more than four 
independent momenta after applying momentum conservation.
Since in collider physics applications the external momenta are
4-dimensional, starting with six-point problems the external momenta
become linearly dependent.
This not only means that there is a constraint on the
kinematics of such processes, but it also leads to new relations
between Feynman integrals.
At one loop, it is well known that this implies that  
six-point integrals can be reduced directly to lower multiplicity
integrals with techniques that are now textbook (see e.g.~\cite{Henn:2014yza}).
While in general it is not true that two-loop six-point integrals 
can be reduced to lower multiplicity integrals, it is still true that the
four-dimensionality of the external kinematics introduces a new class of
relations between these integrals.
The second property is related to the number of propagators that two-loop
six-point topologies can have.
Indeed, for the first time we find two-loop topologies with more than eight propagators,
and the number of integration variables in the four-dimensional limit (that
is, when the loop momenta are taken to be four dimensional) is 
fewer than the number of propagators. 
At the one-loop level, this occurs for the pentagon,
and signals a reduction identity that is valid up to $\mathcal{O}(\epsilon)$
corrections in the dimensional
regulator~\cite{Bern:1993kr}. 
It is thus natural to ask if the same happens at two-loops, that is
if nine-propagator integrals can be arranged to decouple in the 
four-dimensional limit at two loops 
(see also~\cite{Bargiela:2024rul} for related work).

In this work, we consider the question of how to tackle the subtleties
of six-point processes in the calculation of Feynman integrals within
the differential equations
method~\cite{Kotikov:1990kg,Kotikov:1991pm,Bern:1993kr,Remiddi:1997ny,Gehrmann:1999as},
and compute the full collection of master integrals for
fully massless planar two-loop six-point processes, such as 4-jet
production.
Firstly, we tackle the problem of manifesting Lorentz-invariance in our
differential equation. To this end, we naturally work in terms of an independent
set of Mandelstam variables. However, at six-point, the set of
linearly-independent Mandelstam invariants themselves satisfy a high-degree
polynomial relation, which we discuss how to treat.
Secondly, we discuss how the linear dependence of the external momenta
induces extra classes of linear relations between the Feynman integrals, and how
we take these into account in the computation.
Thirdly, we adapt standard techniques for the construction of a basis of master
integrals that satisfies a so-called ``canonical'' differential
equation~\cite{Henn:2013pwa} to the case of six-point kinematics.
An important step in this construction is to find the general solution of the
differential equation in the $\epsilon \rightarrow 0$ limit, which can be used
to change basis to a canonical form.
Two problems arise in this case. First, we must determine the differential
equation itself in the $\epsilon \rightarrow 0$ limit, which is a highly
non-trivial task due to the large number of variables in the problem.
Second, we must then solve this multivariate differential equation subject to
the constraint that the external momenta are linearly dependent.
To tackle these difficulties we work with finite-field
techniques~\cite{vonManteuffel:2014ixa,Peraro:2016wsq} for the
construction of the differential equations. We
propose to exploit the analytic structure of the change of basis
matrix to construct an ansatz for the $\epsilon \rightarrow 0$ limit
of the differential equation, which we constrain using finite-field
reductions from \textsc{Kira}~\cite{Klappert:2020nbg}.  Our ansatz
both has a dramatically limited number of free parameters and also
trivializes the solution of the $\epsilon \rightarrow 0$ limit of the
differential equation, allowing us to efficiently construct a pure
basis for our space of six-point master integrals.
Finally, we show that it is indeed possible to choose a basis of integrals
such that the nine-propagator topologies decouple in the four-dimensional limit.
We therefore conclude that, for physical applications where external
momenta are four dimensional, we can side-step the computation of the most 
complicated six-point topologies.

The paper is structured as follows: in section 2, we discuss
four-dimensional six-point kinematics and how to construct
differential operators which respect the four-dimensionality of the
external momenta. In section 3, we define the six-point integral
families that we study, and show that it is sufficient to study only 
eight-propagator topologies. In section 4, we discuss our construction of a
canonical basis, highlighting our analytic reconstruction method for
the canonicalization. In section 5, we solve the differential
equations numerically with the
\textsc{AMFlow}~\cite{Liu:2017jxz,Liu:2021wks} and
\textsc{DiffExp}~\cite{Hidding:2020ytt} codes, highlighting the
subtleties of their applications to four-dimensional kinematics.

\section{Six-point Kinematics in Four Dimensions}
\label{sec:KinematicsNotation}

We consider the scattering of six massless particles, with
momenta $p_i$, $i = 1, \ldots, 6$, which satisfy momentum
conservation $\sum_{i = 1}^6 p_i = 0$ and the on-shell conditions
$p_i^2 = 0$.
A set of 9 Mandelstam invariants which solve the momentum conservation and
the on-shell constraints can be chosen as
\begin{equation}
   \vec{s} = \{s_{12}, s_{23}, s_{34}, s_{45}, s_{56}, s_{16}, s_{123}, s_{234}, s_{345}\}\,,
   \label{eq:CyclicMandelstams}
\end{equation}
where $s_{ij}=(p_i+p_j)^2$ and $s_{ijk}=(p_i+p_j+p_k)^2$.

An important fact that underlies many of the novel features of the calculation
in this paper is that we impose that the $p_i$ are strictly four-dimensional vectors.
Given that the integrals we compute are Lorentz invariant functions, 
it is convenient to consider them as functions of the Mandelstam invariants,
which in turn implies that we must translate the
four-dimensionality constraints to the Mandelstam variables.
This can be achieved by using the Gram determinant of a set
of momenta, defined as 
\begin{equation}
  G
  \left(
    \begin{array}{ccc} \!u_1\! & \ldots & \!u_n\! \\
        \!v_1\! & \ldots & \!v_n\!
    \end{array} 
   \right) = \det(2 u_i \cdot v_j)\,,
\end{equation}
where, in this paper, the $u_i$ and $v_i$ are vectors in Minkowski space.
We will frequently make use of the case where $u_i = v_i = p_i$,
and for convenience we define the short-hand notation
\begin{equation}\label{eq:oneLineGram}
G(p_{i_1} p_{i_2} p_{i_3} p_{i_4}) = 
  G\left(
  \begin{array}{cccc}
    \!p_{i_1}\! & \!p_{i_2}\! & \!p_{i_3}\! & \!p_{i_4}\! \\
    \!p_{i_1}\! & \!p_{i_2}\! & \!p_{i_3}\! & \!p_{i_4}\! 
  \end{array}
    \right)\,.
\end{equation}
Furthermore, to keep the notation light, we will often simply 
indicate the indices of the vectors entering the Gram determinant and
write
\begin{equation}\label{eq:twoLineGram}
    G\left(
  \begin{array}{cccc}
    \!i_1\! & \!i_2\! & \!i_3\! & \!i_4\! \\
    \!j_1\! & \!j_2\! & \!j_3\! & \!j_4\! 
  \end{array}
    \right)=
    G\left(
  \begin{array}{cccc}
    \!p_{i_1}\! & \!p_{i_2}\! & \!p_{i_3}\! & \!p_{i_4}\!\\
    \!p_{j_1}\! & \!p_{j_2}\! & \!p_{j_3}\! & p_{j_4}\!
  \end{array}
  \right)\,,
\end{equation}
and analogously for Eq.~\eqref{eq:oneLineGram}.

The consequences of the four dimensionality of the external momenta on
the Lorentz-invariant Mandelstam variables in $\vec s$ (see
Eq.~\eqref{eq:CyclicMandelstams}) can be expressed neatly in terms of
Gram determinants.  Specifically, as any five four-dimensional momenta
are linearly dependent, one has that
\begin{equation}
  \Delta_6 = G(1 2 3 4 5) = 0\,.
  \label{eq:SixPointGramConstraint}
\end{equation}
That is, the four-dimensionality of the external momenta induces a single
constraint on the 9 Mandelstam variables, which in practice means that there
are only 8 independent Mandelstam variables. While $G(1 2 3 4 5)$ is
a degree-five polynomial in the variables in $\vec s$, it is quadratic in each
of them. This implies that describing the phase space in terms of 
an independent set of Mandelstams requires the introduction of
a square root to solve Eq.~\eqref{eq:SixPointGramConstraint}. 
As there is no manifestly symmetric way to solve the six-point Gram
constraint, we arbitrarily choose
to solve $\Delta_6 = 0$ with respect to $s_{16}$. 
Given this choice, an
important quantity that will feature in our calculation is the discriminant of
the quadratic equation $\Delta_6 = 0$ with respect to $s_{16}$. 
This is given by
\begin{equation}
  \mathrm{Discriminant}_{s_{16}}( \Delta _6 ) = 4\,G(1234) G(3456)\,,
\end{equation}
where we made explicit the interesting fact that the discriminant 
factorizes into a product of (five-point) Gram determinants.
We note nevertheless that throughout this paper we will strive to avoid
explicitly solving the constraint of Eq.~\eqref{eq:SixPointGramConstraint}
and instead present expressions that depend on the 9 invariants
in $\vec s$. These expressions are valid modulo $G(1 2 3 4 5)=0$
and are thus not unique, but have the advantage of keeping a square root
implicit. When evaluated in four-dimensional kinematics this constraint
is automatically satisfied and the result is therefore unique.

The four dimensionality of the external momenta also has consequences
on the five-point Gram determinants $G(i_1i_2i_3i_4)$.  Indeed, taking
into account trivial determinant relations, there are five different
$G(i_1i_2i_3i_4)$ one can consider, but the four-dimensionality of the
external kinematics implies that these five determinants are dependent
on each other.  The Lorentz invariant consequences of this observation
can be made explicit using $\Delta_6 = 0$ together with
Sylvester's determinant identity, which leads to
\begin{equation}
  G(i_1 i_2 i_3 i_4)
  G(j_1 j_2 j_3 j_4)
  = G\left( \begin{array}{cccc}
\!i_1\! & \!i_2\! & \!i_3\! & \!i_4\! \\
\!j_1\! & \!j_2\! & \!j_3\! & \!j_4\!
  \end{array} \right)^2\,.
\label{eq:GramDeterminantRelations}
\end{equation}
While this relation might seem unremarkable, it has subtle consequences
when taking square-roots of Gram determinants. Indeed,
the latter are ubiquitous in parametrising phase space or in
defining pure bases of master integrals and are thus natural quantities
to consider. 
Taking the square-root of the relation in Eq.~\eqref{eq:GramDeterminantRelations}
implies that the use of multiple such square roots leads to a multiple
cover of phase space. While this is not problematic in principle, the 
multiple cover is entirely unnecessary and can be avoided. Specifically, we
define a canonical Gram determinant square root as
\begin{equation}\label{eq:delta5}
  \sqrt{\Delta_5} = \sqrt{G(1 2 3 4)}\,.
\end{equation}
We then define all other five-point square roots in terms of this
canonical choice as
\begin{equation}
    \sqrt{\Delta^{(5)}_{i_1 i_2 i_3 i_4}} = \frac{G\left( \begin{array}{cccc}
\!i_1\! & \!i_2\! & \!i_3\! & \!i_4\! \\
\!1\! & \!2\! & \!3\! & \!4\!
  \end{array} \right)}{\sqrt{\Delta_5}}\,,
\label{eq:ExtraGramRootDefinition}
\end{equation}
which is a rational relation without any branch ambiguities.
Given the relation \eqref{eq:GramDeterminantRelations} between Gram determinants, 
one has that 
\begin{equation}
    \left(\sqrt{\Delta^{(5)}_{i_1 i_2 i_3 i_4}}\right)^2 = G(i_1 i_2 i_3 i_4)\,,
\end{equation}
but importantly only a single algebraic branch cut has been introduced.

Besides five-point Gram determinants, we will also encounter
the (square-roots of) three-point Gram determinants
\begin{equation}\label{eq:lambdaDef}
  G(p_{i_1i_2},p_{i_3i_4})=-\lambda(p_{i_1i_2}^2,p_{i_3i_4}^2,p_{i_5i_6}^2)
\end{equation}
where all $i_k$ are distinct, we defined $p_{ab}=p_{a}+p_{b}$,
and $\lambda(a,b,c)=a^2+b^2+c^2-2ab-2ac-2bc$ is the K\"all\'en function.

\subsection{Differential Operators}

Our goal is to compute six-point Feynman integrals by solving
the differential equations they satisfy. In order to obtain
the differential equations, we must first
construct a set of differential operators which respect the $\Delta_6 = 0$
constraint. 
We treat this constraint as any other constraint that is imposed
on the kinematics (such as momentum conservation or the on-shell condition),
and note that the various conditions that are placed
on the kinematics impose linear constraints on the basic differentials 
that we use to express the total differential. 
We note that we apply this approach in a fully numerical determination of the
differential equation. Therefore, the constraints can be solved numerically,
phase-space point by phase-space point, which greatly simplifies the approach.

We begin by considering a large, redundant set of
variables: both the components of the external momenta $p_i^\mu$, and the cyclic
Mandelstams $\vec{s}$. As an operator, we can express the total derivative as
\begin{equation}\label{eq:totalDiffGen}
      \mathrm{d} = \sum_{i} \mathrm{d} s_{i, i+1} \frac{\partial}{\partial s_{i, i+1}} + \sum_{i} \mathrm{d} s_{i, i+1, i+2} \frac{\partial}{\partial s_{i, i+1, i+2}} + \sum_{i = 1}^6 \sum_{\mu=0}^3 \mathrm{d}p_i^\mu \frac{\partial}{\partial p_i^\mu}\,,
\end{equation}
where we have explicitly implemented the four-dimensionality of the momenta in
that the sum over $\mu$ runs only over four components. For practical reasons it
is more convenient if all differential operators are Lorentz invariant,
so we introduce
\begin{equation}
  v_i = \sum_{j=1}^4\mathcal{G}(1234)^{-1}_{ij} p_j\,,
\end{equation}
which satisfies
\begin{equation}
  g^{\mu \nu} = \sum_{j=1}^4 v_j^\mu p_j^\nu\,,
\end{equation}
where $g^{\mu \nu}$ is the four-dimensional metric. 
This allows us to write
\begin{equation}
      \sum_{i = 1}^6 \sum_{\mu=0}^3 \mathrm{d}p_i^\mu \frac{\partial}{\partial p_i^\mu} = \sum_{i = 1}^6 \sum_{j=1}^4 (v_{j} \cdot \mathrm{d} p_i) \left(p_j \cdot \frac{\partial}{\partial p_i}\right)\,.
\end{equation}
Substituting this relation into Eq.~\eqref{eq:totalDiffGen}, 
the total derivative is written in terms of manifestly Lorentz
invariant operators.
We can then impose that the total differential is consistent with 
the definition of the Mandelstam invariants,
\begin{equation}
  \mathrm{d} \left[s_{i, i+1} - (p_{i} + p_{i+1})^2\right] = 0, 
  \qquad \mathrm{d} 
  \left[s_{i, i+1, i+2} - (p_{i} + p_{i+1} + p_{i+2})^2\right] = 0\,,
  \label{eq:DifferentialDefinitionConstraints}
\end{equation}
and with the on-shell condition and momentum conservation constraints
\begin{equation}
  \mathrm{d}[p_i^2] = 0 \quad {i = 1, \ldots, 6}, \qquad 
  p_j \cdot \mathrm{d}\left[\sum_{i = 1}^6 p_i\right]  = 0 \quad {j=1, 
  \ldots, 4}.
  \label{eq:DifferentialMomConsAndOnShellConstraints}
\end{equation}
Each of these constraints can be interpreted as a (linear) constraint on the
$\mathrm{d}s_{i, i+1}$, $\mathrm{d}s_{i, i+1, i+2}$ and $v_j \cdot \mathrm{d}
p_i$, whose coefficients are rational in Mandelstam invariants. Therefore, for each
phase-space point, we sample Eqs.~\eqref{eq:DifferentialDefinitionConstraints} and
\eqref{eq:DifferentialMomConsAndOnShellConstraints} and solve for a basis of the
differentials.
This solution is not unique, but we find that the degrees of freedom correspond
to differential operators which annihilate Lorentz invariant functions and we
are therefore allowed to set their associated unknowns to zero.
Having done this, we find, as expected, that only 8 such differentials are
independent, and we choose them to be the differentials corresponding to our
independent set of Mandelstam invariants (to be consistent with the discussion
below
Eq.~\eqref{eq:SixPointGramConstraint}, all Mandelstam invariants in $\vec s$ 
apart from $s_{16}$). 
This provides a practical solution to computing the total derivatives
of Feynman integrals while imposing all constraints, including the four-dimensionality
of the external kinematics.

The analytic form of the total differential is not particularly simple, 
but this is easily bypassed by constructing the operator numerically at
each phase-space point.
A key element of this setup is to be able to generate a
configuration of 9 Mandelstam variables which satisfy the constraint 
$\Delta_6 =0$. As noted above, this constraint is
quadratic. It is therefore simple to apply the finite-field veto strategy
introduced in \cite{Abreu:2020jxa}, where random values of 8 of the Mandelstams
are chosen until a solution to $\Delta_6 = 0$ is found within the finite 
field.

\section{Integral Topologies}
\label{sec:IntegralDefinitions}

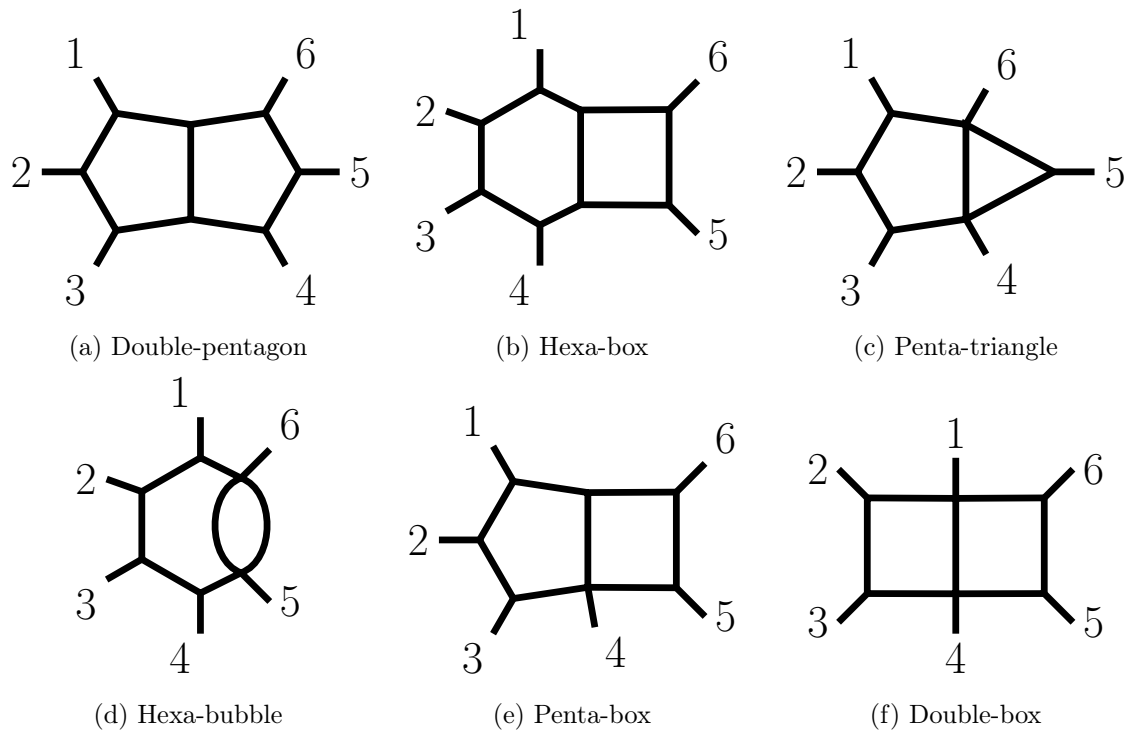
\begin{figure}[t]
  \centering

    \begin{subfigure}[b]{0.32\textwidth}
         \centering

    \begin{tikzpicture}[scale=0.9]
    \tikzmath{\width=0.9; \r1=1; \midpoint=1.2; \legL=0.6;}

    % Central vertex placement
    \tikzmath{\xVCu={0.5*\midpoint}; \yVCu= {0.7*\r1};}
    \tikzmath{\xVCd={0.5*\midpoint}; \yVCd= {-0.7*\r1};}

    \tikzmath{\xV1={\r1*cos(120)}; \yV1={\r1*sin(120)};}
    \tikzmath{\xV2={\r1*cos(180)}; \yV2={\r1 * sin(180)};}
    \tikzmath{\xV3={\r1*cos(240)}; \yV3={\r1 * sin(240)};}
    \tikzmath{\xV4={\midpoint + \r1*cos(300)}; \yV4={\r1*sin(300)};}
    \tikzmath{\xV5={\midpoint + \r1*cos(0)}; \yV5={\r1*sin(0)};}
    \tikzmath{\xV6={\midpoint + \r1*cos(60)}; \yV6={\r1*sin(60)};}

    % propagators
    % central -- p1  -- p2 -- p3 -- central -- p4 -- p5 -- p6 -- central -- central
    \draw [line width=\width mm] (\xVCu, \yVCu) -- ( \xV1, \yV1 ) -- ( \xV2 , \yV2 ) -- ( \xV3 , \yV3 ) -- (\xVCd, \yVCd) -- (\xV4, \yV4) -- (\xV5, \yV5) -- (\xV6, \yV6) -- (\xVCu, \yVCu) -- (\xVCd, \yVCd);

    % Right Legs
    \draw [line width=\width mm] (\xV4, \yV4) -- ({\xV4 + \legL*cos(300)}, {\yV4 + \legL*sin(300)});
    \node [below right,font={\LARGE}] at ({\xV4 + \legL*cos(300)}, {\yV4 + \legL*sin(300)})  {$4$};

    \draw [line width=\width mm] (\xV5, \yV5) -- ({\xV5 + \legL}, \yV5);
    \node [right,font={\LARGE}] at ({\xV5 + \legL}, \yV5)  {$5$};

    \draw [line width=\width mm] (\xV6, \yV6)  -- ({\xV6 + \legL*cos(60)}, {\yV6 + \legL*sin(60)});
    \node [above right,font={\LARGE}] at ({\xV6 + \legL*cos(60)}, {\yV6 + \legL*sin(60)})   {$6$};

    % Left Legs
    \draw [line width=\width mm] (\xV1, \yV1)  -- ({\xV1 + \legL*cos(120)}, {\yV1 + \legL*sin(120)});
    \node [above left,font={\LARGE}] at ({\xV1 + \legL*cos(120)}, {\yV1 + \legL*sin(120)})  {$1$};

    \draw [line width=\width mm] (\xV2, \yV2) -- (\xV2 - \legL, \yV2);
    \node [left,font={\LARGE}] at  (\xV2 - \legL, \yV2) {$2$};

    \draw [line width=\width mm] (\xV3, \yV3) -- ({\xV3 + \legL*cos(240)}, {\yV3 + \legL*sin(240)});
    \node [below left,font={\LARGE}] at ({\xV3 + \legL*cos(240)}, {\yV3 + \legL*sin(240)}) {$3$};

    \end{tikzpicture} 

         \caption{Double-pentagon}
         \label{fig:dpTopology}
     \end{subfigure}
    \begin{subfigure}[b]{0.32\textwidth}
         \centering

    \begin{tikzpicture}[scale=0.9]
    \tikzmath{\width=0.9; \r1=1; \midpoint=1.2; \legL=0.6;}

    % Central vertex placement
    \tikzmath{\xVCu={0.5*\midpoint}; \yVCu= {0.7*\r1};}
    \tikzmath{\xVCd={0.5*\midpoint}; \yVCd= {-0.7*\r1};}

    \tikzmath{\xV1={\r1*cos(90)}; \yV1={\r1*sin(90)};}
    \tikzmath{\xV2={\r1*cos(150)}; \yV2={\r1 * sin(150)};}
    \tikzmath{\xV3={\r1*cos(210)}; \yV3={\r1 * sin(210)};}
    \tikzmath{\xV4={\r1*cos(270)}; \yV4={\r1 * sin(270)};}
    \tikzmath{\xV5={\midpoint + \r1*cos(315)}; \yV5={\r1*sin(315)};}
    \tikzmath{\xV6={\midpoint + \r1*cos(45)}; \yV6={\r1*sin(45)};}

    % propagators
    % central -- p1  -- p2 -- p3 -- p4 -- central -- p5 -- p6 -- central -- central
    \draw [line width=\width mm] (\xVCu, \yVCu) -- ( \xV1, \yV1 ) -- ( \xV2 , \yV2 ) -- ( \xV3 , \yV3 ) -- (\xV4, \yV4) -- (\xVCd, \yVCd)  -- (\xV5, \yV5) -- (\xV6, \yV6) -- (\xVCu, \yVCu) -- (\xVCd, \yVCd);

    % Right Legs

    \draw [line width=\width mm] (\xV5, \yV5) -- ({\xV5 + cos(315)*\legL}, {\yV5 + sin(315)*\legL});
    \node [right,font={\LARGE}] at ({\xV5 + cos(315)*\legL}, {\yV5 + sin(315)*\legL})  {$5$};

    \draw [line width=\width mm] (\xV6, \yV6)  -- ({\xV6 + \legL*cos(45)}, {\yV6 + \legL*sin(45)});
    \node [above right,font={\LARGE}] at ({\xV6 + \legL*cos(45)}, {\yV6 + \legL*sin(45)})   {$6$};

    % Left Legs

    \draw [line width=\width mm] (\xV1, \yV1)  -- ({\xV1 + \legL*cos(90)}, {\yV1 + \legL*sin(90)});
    \node [above left,font={\LARGE}] at ({\xV1 + \legL*cos(90)}, {\yV1 + \legL*sin(90)})  {$1$};

    \draw [line width=\width mm] (\xV2, \yV2) -- ({\xV2 + cos(150)*\legL}, {\yV2 + sin(162)*\legL});
    \node [left,font={\LARGE}] at  ({\xV2 + cos(150)*\legL}, {\yV2 + sin(162)*\legL}) {$2$};

    \draw [line width=\width mm] (\xV3, \yV3) -- ({\xV3 + \legL*cos(210)}, {\yV3 + \legL*sin(210)});
    \node [below left,font={\LARGE}] at ({\xV3 + \legL*cos(210)}, {\yV3 + \legL*sin(210)}) {$3$};

    \draw [line width=\width mm] (\xV4, \yV4) -- ({\xV4 + \legL*cos(270)}, {\yV4 + \legL*sin(270)});
    \node [below left,font={\LARGE}] at ({\xV4 + \legL*cos(270)}, {\yV4 + \legL*sin(270)})  {$4$};

    \end{tikzpicture} 

         \caption{Hexa-box}
         \label{fig:hbTopology}
     \end{subfigure}
    \begin{subfigure}[b]{0.32\textwidth}
         \centering

    \begin{tikzpicture}[scale=0.9]
    \tikzmath{\width=0.9; \r1=1; \midpoint=1.2; \legL=0.6;}

    % Central vertex placement
    \tikzmath{\xVCu={0.5*\midpoint}; \yVCu= {0.7*\r1};}
    \tikzmath{\xVCd={0.5*\midpoint}; \yVCd= {-0.7*\r1};}

    \tikzmath{\xV1={\r1*cos(120)}; \yV1={\r1*sin(120)};}
    \tikzmath{\xV2={\r1*cos(180)}; \yV2={\r1 * sin(180)};}
    \tikzmath{\xV3={\r1*cos(240)}; \yV3={\r1 * sin(240)};}
    \tikzmath{\xV4=\xVCd; \yV4=\yVCd;}
    \tikzmath{\xV5={0.75*\midpoint + \r1*cos(0)}; \yV5={\r1*sin(0)};}
    \tikzmath{\xV6=\xVCu; \yV6=\yVCu;}

    % propagators
    % central -- p1  -- p2 -- p3 -- central -- p4 -- p5 -- p6 -- central -- central
    \draw [line width=\width mm] (\xVCu, \yVCu) -- ( \xV1, \yV1 ) -- ( \xV2 , \yV2 ) -- ( \xV3 , \yV3 ) -- (\xVCd, \yVCd) -- (\xV4, \yV4) -- (\xV5, \yV5) -- (\xV6, \yV6) -- (\xVCu, \yVCu) -- (\xVCd, \yVCd);

    % Right Legs
    \draw [line width=\width mm] (\xV4, \yV4) -- ({\xV4 + \legL*cos(300)}, {\yV4 + \legL*sin(300)});
    \node [below right,font={\LARGE}] at ({\xV4 + \legL*cos(300)}, {\yV4 + \legL*sin(300)})  {$4$};

    \draw [line width=\width mm] (\xV5, \yV5) -- ({\xV5 + \legL}, \yV5);
    \node [right,font={\LARGE}] at ({\xV5 + \legL}, \yV5)  {$5$};

    \draw [line width=\width mm] (\xV6, \yV6)  -- ({\xV6 + \legL*cos(60)}, {\yV6 + \legL*sin(60)});
    \node [above right,font={\LARGE}] at ({\xV6 + \legL*cos(60)}, {\yV6 + \legL*sin(60)})   {$6$};

    % Left Legs
    \draw [line width=\width mm] (\xV1, \yV1)  -- ({\xV1 + \legL*cos(120)}, {\yV1 + \legL*sin(120)});
    \node [above left,font={\LARGE}] at ({\xV1 + \legL*cos(120)}, {\yV1 + \legL*sin(120)})  {$1$};

    \draw [line width=\width mm] (\xV2, \yV2) -- (\xV2 - \legL, \yV2);
    \node [left,font={\LARGE}] at  (\xV2 - \legL, \yV2) {$2$};

    \draw [line width=\width mm] (\xV3, \yV3) -- ({\xV3 + \legL*cos(240)}, {\yV3 + \legL*sin(240)});
    \node [below left,font={\LARGE}] at ({\xV3 + \legL*cos(240)}, {\yV3 + \legL*sin(240)}) {$3$};

    \end{tikzpicture} 

         \caption{Penta-triangle}
         \label{fig:ptTopology}
     \end{subfigure}

    \begin{subfigure}[b]{0.32\textwidth}
         \centering

    \begin{tikzpicture}[scale=0.9]
    \tikzmath{\width=0.9; \r1=1; \midpoint=1.2; \legL=0.6;}

    % Central vertex placement
    \tikzmath{\xVCu={0.5*\midpoint}; \yVCu= {0.7*\r1};}
    \tikzmath{\xVCd={0.5*\midpoint}; \yVCd= {-0.7*\r1};}

    \tikzmath{\xV1={\r1*cos(90)}; \yV1={\r1*sin(90)};}
    \tikzmath{\xV2={\r1*cos(150)}; \yV2={\r1 * sin(150)};}
    \tikzmath{\xV3={\r1*cos(210)}; \yV3={\r1 * sin(210)};}
    \tikzmath{\xV4={\r1*cos(270)}; \yV4={\r1 * sin(270)};}

    \tikzmath{\xV5=\xVCd; \yV5=\yVCd;}
    \tikzmath{\xV6=\xVCu; \yV6=\yVCu;}

    % propagators
    % central -- p1  -- p2 -- p3 -- p4 -- central -- p5 -- p6 -- central -- central
    \draw [line width=\width mm] (\xVCu, \yVCu) -- ( \xV1, \yV1 ) -- ( \xV2 , \yV2 ) -- ( \xV3 , \yV3 ) -- (\xV4, \yV4) -- (\xVCd, \yVCd)  -- (\xV5, \yV5) to[out=20,in=-20] (\xV6, \yV6) -- (\xVCu, \yVCu) to[out=200,in=160] (\xVCd, \yVCd);

    % Right Legs

    \draw [line width=\width mm] (\xV5, \yV5) -- ({\xV5 + cos(315)*\legL}, {\yV5 + sin(315)*\legL});
    \node [right,font={\LARGE}] at ({\xV5 + cos(315)*\legL}, {\yV5 + sin(315)*\legL})  {$5$};

    \draw [line width=\width mm] (\xV6, \yV6)  -- ({\xV6 + \legL*cos(45)}, {\yV6 + \legL*sin(45)});
    \node [above right,font={\LARGE}] at ({\xV6 + \legL*cos(45)}, {\yV6 + \legL*sin(45)})   {$6$};

    % Left Legs

    \draw [line width=\width mm] (\xV1, \yV1)  -- ({\xV1 + \legL*cos(90)}, {\yV1 + \legL*sin(90)});
    \node [above left,font={\LARGE}] at ({\xV1 + \legL*cos(90)}, {\yV1 + \legL*sin(90)})  {$1$};

    \draw [line width=\width mm] (\xV2, \yV2) -- ({\xV2 + cos(150)*\legL}, {\yV2 + sin(162)*\legL});
    \node [left,font={\LARGE}] at  ({\xV2 + cos(150)*\legL}, {\yV2 + sin(162)*\legL}) {$2$};

    \draw [line width=\width mm] (\xV3, \yV3) -- ({\xV3 + \legL*cos(210)}, {\yV3 + \legL*sin(210)});
    \node [below left,font={\LARGE}] at ({\xV3 + \legL*cos(210)}, {\yV3 + \legL*sin(210)}) {$3$};

    \draw [line width=\width mm] (\xV4, \yV4) -- ({\xV4 + \legL*cos(270)}, {\yV4 + \legL*sin(270)});
    \node [below left,font={\LARGE}] at ({\xV4 + \legL*cos(270)}, {\yV4 + \legL*sin(270)})  {$4$};

    \end{tikzpicture} 

         \caption{Hexa-bubble}
         \label{fig:hbubTopology}
     \end{subfigure}
    \begin{subfigure}[b]{0.32\textwidth}
         \centering

    \begin{tikzpicture}[scale=0.9]
    \tikzmath{\width=0.9; \r1=1; \midpoint=1.2; \legL=0.6;}

    % Central vertex placement
    \tikzmath{\xVCu={0.5*\midpoint}; \yVCu= {0.7*\r1};}
    \tikzmath{\xVCd={0.5*\midpoint}; \yVCd= {-0.7*\r1};}

    \tikzmath{\xV1={\r1*cos(120)}; \yV1={\r1*sin(120)};}
    \tikzmath{\xV2={\r1*cos(180)}; \yV2={\r1 * sin(180)};}
    \tikzmath{\xV3={\r1*cos(240)}; \yV3={\r1 * sin(240)};}
    \tikzmath{\xV5 = \midpoint + \r1*cos(315); \yV5 = \r1*sin(315);}
    \tikzmath{\xV6 = \midpoint + \r1*cos(45); \yV6 = \r1*sin(45);}

    \tikzmath{\xV4=\xVCd; \yV4=\yVCd;}

    % propagators
    % central -- p1  -- p2 -- p3 -- central -- p4 -- p5 -- p6 -- central -- central
    \draw [line width=\width mm] (\xVCu, \yVCu) -- ( \xV1, \yV1 ) -- ( \xV2 , \yV2 ) -- ( \xV3 , \yV3 ) -- (\xVCd, \yVCd) -- (\xV4, \yV4) -- (\xV5, \yV5) -- (\xV6, \yV6) -- (\xVCu, \yVCu) -- (\xVCd, \yVCd);

    % Bottom Leg
    \draw [line width=\width mm] (\xV4, \yV4) -- ({\xV4 + \legL*cos(280)}, {\yV4 + \legL*sin(280)});
    \node [below right,font={\LARGE}] at ({\xV4 + \legL*cos(280)}, {\yV4 + \legL*sin(280)})  {$4$};

    % Right Legs
    \draw [line width=\width mm] (\xV5, \yV5) -- ({\xV5 + \legL*cos(315)}, {\yV5 + \legL*sin(315)});
    \node [right,font={\LARGE}] at ({\xV5 + \legL*cos(315)}, {\yV5 + \legL*sin(315)})  {$5$};

    \draw [line width=\width mm] (\xV6, \yV6)  -- ({\xV6 + \legL*cos(45)}, {\yV6 + \legL*sin(45)});
    \node [above right,font={\LARGE}] at ({\xV6 + \legL*cos(45)}, {\yV6 + \legL*sin(45)})   {$6$};

    % Left Legs
    \draw [line width=\width mm] (\xV1, \yV1)  -- ({\xV1 + \legL*cos(120)}, {\yV1 + \legL*sin(120)});
    \node [above left,font={\LARGE}] at ({\xV1 + \legL*cos(120)}, {\yV1 + \legL*sin(120)})  {$1$};

    \draw [line width=\width mm] (\xV2, \yV2) -- (\xV2 - \legL, \yV2);
    \node [left,font={\LARGE}] at  (\xV2 - \legL, \yV2) {$2$};

    \draw [line width=\width mm] (\xV3, \yV3) -- ({\xV3 + \legL*cos(240)}, {\yV3 + \legL*sin(240)});
    \node [below left,font={\LARGE}] at ({\xV3 + \legL*cos(240)}, {\yV3 + \legL*sin(240)}) {$3$};

    \end{tikzpicture} 

         \caption{Penta-box}
         \label{fig:pbTopology}
     \end{subfigure}
    \begin{subfigure}[b]{0.32\textwidth}
         \centering

    \begin{tikzpicture}[scale=0.9]
    \tikzmath{\width=0.9; \r1=1; \midpoint=1.2; \legL = 0.6;}
    % propagators

    \tikzmath{\xV1 = \r1 * cos(135); \yV1 = \r1*sin(135);}
    \tikzmath{\xV2 = \r1 * cos(225); \yV2 = \r1*sin(225);}
    \tikzmath{\xV3 = 0.5*\midpoint; \yV3 = -0.7*\r1;}
    \tikzmath{\xV4 = \midpoint + \r1*cos(315); \yV4 = \r1*sin(315);}
    \tikzmath{\xV5 = \midpoint + \r1*cos(45); \yV5 = \r1*sin(45);}
    \tikzmath{\xV6 = 0.5*\midpoint; \yV6 = 0.7*\r1;}

    % central (p6) -- p1  -- p2 -- central (p3) -- p4 -- p5 -- central (p6) -- central (p3)

    \draw [line width=\width mm] ({0.5*\midpoint}, {0.7*\r1}) -- ( {\xV1} , {\yV1} ) -- ( {\xV2} , {\yV2} ) -- ({0.5*\midpoint}, {-0.7 \r1}) -- (\xV4, \yV4) -- (\xV5, \yV5) -- (\xV6, \yV6) -- (\xV3, \yV3);

    % Left Legs
    \draw [line width=\width mm] (\xV1, \yV1) -- ({\xV1 + \legL*cos(135)}, {\yV1 + \legL*sin(135)});
    \node [left,font={\LARGE}] at ({\xV1 + \legL*cos(135)}, {\yV1 + \legL*sin(135)})  {$2$};

    \draw [line width=\width mm] (\xV2, \yV2) -- ({\xV2 + \legL*cos(225)}, {\yV2 + \legL*sin(225)});
    \node [left,font={\LARGE}] at  ({\xV2 + \legL*cos(225)}, {\yV2 + \legL*sin(225)}) {$3$};

    % Bottom Leg
    \draw [line width=\width mm] (\xV3, \yV3) -- (\xV3, {\yV3 - \legL});
    \node [below,font={\LARGE}] at (\xV3, {\yV3 - \legL}) {$4$};

    % Right Legs
    \draw [line width=\width mm] (\xV5, \yV5) -- ({\xV5 + \legL*cos(45)}, {\yV5 + \legL*sin(45)});
    \node [right,font={\LARGE}] at ({\xV5 + \legL*cos(45)}, {\yV5 + \legL*sin(45)})  {$6$};

    \draw [line width=\width mm] (\xV4, \yV4) -- ({\xV4 + \legL*cos(315)}, {\yV4 + \legL*sin(315)});
    \node [right,font={\LARGE}] at ({\xV4 + \legL*cos(315)}, {\yV4 + \legL*sin(315)})  {$5$};

    % Top leg
    \draw [line width=\width mm] (\xV6, \yV6)  -- (\xV6, {\yV6 + \legL});
    \node [above,font={\LARGE}] at (\xV6, {\yV6 + \legL})   {$1$};

    \end{tikzpicture} 

         \caption{Double-box}
         \label{fig:dbTopology}
     \end{subfigure}
     
     \caption{The two-loop six-point integral topologies considered in this
       work. All two-loop six-point planar integrals can be reduced to these
       topologies by integration by parts relations.}
  \label{fig:SixPointIntegralTopologies}

\end{figure}

\begin{figure}[t]
  \centering
  \includegraphics[width=.5\textwidth]{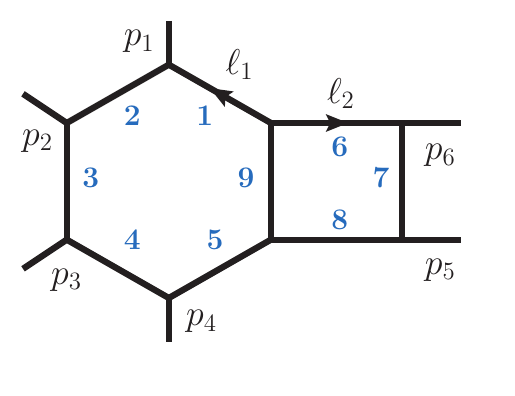}
  \caption{Hexa-box topology. The blue labels give the index
  of the propagators according to the conventions 
  of Eq.~\eqref{eq:propagators}.}
  \label{fig:hboxParam}
\end{figure}

It is well known that Feynman integrals form a vector space under
so-called ``integration by parts'' (IBP)
relations~\cite{Tkachov:1981wb,Chetyrkin:1981qh}. In this work, we
construct a basis of this vector space, the so called ``master
integrals'' (MIs). Many of the integrals under consideration here have
already been treated in earlier works. See, for example, the set of
five-point one-mass planar integrals~\cite{Abreu:2020jxa}.
The two-loop six-point topologies that have a non-zero number of
master integrals are depicted in
Fig.~\ref{fig:SixPointIntegralTopologies}.  As we will discuss in
section~\ref{sec:9PropagatorIrrelevance}, for four-dimensional gauge
theories, the computation of the double-pentagon topology is unnecessary
(as well as, as we will show, the hexa-box integral itself).  It is therefore
clear that all such two-loop six-point integrals can be organized into
a single top topology, the hexa-box in Fig.~\ref{fig:hboxParam}.
Consequently, we can write all the integrals  we will consider
in terms of the topology
\begin{equation}\label{eq:topologyDef}
    I[\vec{\nu}] = \int \frac{\mathrm{d}^D\ell_1}{i \pi^{D/2}} \frac{\mathrm{d}^D\ell_2}{i \pi^{D/2}} \frac{\rho_{10}^{-\nu_{10}} \cdots \rho_{10}^{-\nu_{13}}}{\rho_1^{\nu_1} \cdots \rho_9^{\nu_9}},
\end{equation}
where the $\nu_i$ are integers, with $\nu_{10},\ldots,\nu_{13}<0$,
and the $\rho_i$ are given by
\begin{align}\label{eq:propagators}
  \begin{split}
  \rho_1 &= \ell_1^2, \rho_2 = (\ell_1 - p_1)^2, \rho_3 = (\ell_1 - p_{12})^2, \rho_4 = (\ell_1 - p_{123})^2, \rho_5 = (\ell_1 - p_{1234})^2 \\
  \rho_6 &= \ell_2^2, \rho_7 = (\ell_2 - p_6)^2, \rho_8 = (\ell_2 - p_{56})^2, \rho_9 = (\ell_1 + \ell_2)^2, \rho_{10} = (\ell_1 - p_{12345})^2, \\
  \rho_{11} &= (\ell_2 - p_{456})^2, \rho_{12} = (\ell_2 - p_{3456})^2, \rho_{13} = (\ell_2 - p_{23456})^2,
  \end{split}
\end{align}
where the external momenta are outgoing and
$p_{i \cdots j} = p_i + \cdots + p_j$. 
For convenience, Fig.~\ref{fig:hboxParam} displays a hexa-box with these
conventions.

The use of four-dimensional external kinematics leads to
additional non-trivial consequences on the number of MIs in the
problem. In particular, further simplifications are observed at the
integrand level, when expressed in terms of Mandelstam
invariants. The constraint on four-dimensional kinematics can be
expressed as
\begin{equation}
  G \left(\begin{array}{ccccc} \!\ell_1\! & \!p_1\! & \!p_2\! & \!p_3\! & \!p_4\! \\ p_5 & \!p_1\! & \!p_2\! & \!p_3\! & \!p_4\! \end{array} \right)  =
  G \left(\begin{array}{ccccc} \!\ell_2\! & \!p_1\! & \!p_2\! & \!p_3\! & \!p_4\! \\ p_5 & \!p_1\! & \!p_2\! & \!p_3\! & \!p_4\! \end{array} \right) = 0\,,
  \label{eq:ISPDependenceConstraint}
\end{equation}
which follows from the linear dependence of the momenta on the bottom
line of the Gram determinant. Naturally, one could consider different
choices of the external momenta in the top line, but 
these give the same constraint given $\Delta_6 = 0$. When expressed in
terms of the propagator variables and ISPs,
Eqs.~\eqref{eq:ISPDependenceConstraint} can be used to eliminate two of the
ISPs, reducing the total number of propagators and ISPs to 11, as
expected for a two-loop integral in the four-dimensional limit.  
In practice, we implement~\eqref{eq:ISPDependenceConstraint} 
by first performing a
numerical IBP reduction of all the Feynman integrals in the problem at
a finite-field point that satisfies $\Delta_6=0$ but
using the full $D$-dimensional kinematics with
\textsc{Kira}~\cite{Klappert:2020nbg}. We then insert
Eqs.~\eqref{eq:ISPDependenceConstraint} in the numerator of each of
the MIs returned by \textsc{Kira}, resulting in additional linear
relations between them. We organise the latter set of equations in the
same format used by \textsc{Kira} for the IBP system and add them to
the system of equations.
We then solve the new system of linear equations (now
containing also the extra relations) and count the number of master
integrals associated to each of the six-point topologies.
The on-shell number of master integrals for each
of the six-point topologies is recorded
in Table~\ref{tab:SixPointMasterIntegralCount}, and agrees
with the results of \cite{Henn:2021cyv}.
\begin{table}[t]
  \centering
  \begin{tabular}{c|c|c|c|c|c}
    Diagram & (b) & (c) & (d) & (e) & (f) \\
    \hline
    Name & Hexa-box & Penta-triangle & Hexa-bubble & Penta-box & Double-box \\
    \hline
    Master Count &1 & 1 & 1 & 3 & 7
  \end{tabular}
  \caption{Master integral count for each of the topologies in
    Fig.~\ref{fig:SixPointIntegralTopologies} apart from the
    double-pentagon. We exclude the double-pentagon, as it does not
    contribute to four-dimensional gauge theory amplitudes, as well as
    all the known integrals related to five-point one-mass planar
    topologies.}

   \label{tab:SixPointMasterIntegralCount}

\end{table}

\subsection{Decoupling of Topologies with 9 Propagators in the $D \rightarrow 4$ Limit}
\label{sec:9PropagatorIrrelevance}

One important consideration when computing high-multiplicity Feynman
integrals that are to be applied in the computation of scattering
amplitudes, is to understand which such integrals contribute to the
four-dimensional limit.  Indeed, at one-loop it is well known that a
basis of Feynman integrals can be chosen which manifests the fact that
the pentagon topology does not contribute~\cite{Bern:1993kr}.
Moreover, in computations of five-point two-loop integrals 
 it is frequently observed that certain master
integrals can be chosen to
only contribute at $\mathcal{O}(\epsilon)$ (see
e.g.~\cite{Abreu:2018rcw}). Nevertheless, these
integrals always formed part of a topology that contained integrals
which did contribute to the finite part of amplitudes, so their full
differential equation was still necessary.

In this section, we argue that at six-point, for the first time, entire
topologies of Feynman integrals decouple in the four-dimensional limit of gauge
theories and therefore do not need to be computed for applications to, say,
collider physics. Specifically, we consider both the so-called ``double-pentagon''
topology and the ``hexa-box'' topology (see Figs.~\ref{fig:dpTopology} and
~\ref{fig:hbTopology}) (note that the other topologies can be embedded
in these two).

\paragraph{Double-Pentagon Topology}
Our aim is to build insertions which span the space of possible numerators, are
irreducible modulo lower topologies and correspond to Feynman integrals that
are manifestly $\mathcal{O}(\epsilon)$.
In the case of the double-pentagon topology, we rely on constraints imposed
by renormalizable power counting. This argumentation is therefore only valid for
a theory that obeys such power counting (such as gauge theories), and would need
to be reconsidered for theories of gravity.
While the power-counting
constraints are simple to implement, the irreducibility and manifestly
$\mathcal{O}(\epsilon)$ constraints require more work. To implement the
irreducibility constraint, we will work on the maximal cut, setting all
propagators to zero. To implement the manifest $\mathcal{O}(\epsilon)$ integral
constraints, we will make use of a class of numerators which give rise to what
has recently been dubbed ``evanescent integrals''~\cite{Gambuti:2023eqh}.

We begin with the double-pentagon topology. Renormalizable power counting tells
us that our numerators should have at most power 5 in each loop momenta, and
total power 8. Note that all such numerators are ultra-violet finite. In the
left and right loop power-counting, the integral scales as a renormalizable
power-counting pentagon which is ultra-violet finite. In the limit where all
loop momenta are large, the scaling of the numerator and measure factor are
beaten by that of the 9 denominators. Therefore, we only need to consider 
infra-red regions in the construction of evanescent integrals.

In order to define infra-red finite numerators, we employ two functions of loop
momenta $\alpha_i$ given by
\begin{equation}
\alpha_1 = \varepsilon_{\mu \nu \rho \sigma}\ell_1^\mu p_1^\nu p_2^\rho p_3^\sigma\, , \quad \textrm{and }  \quad \alpha_2 = \varepsilon_{\mu \nu \rho \sigma} \ell_2^\mu p_4^\nu p_5^\rho p_6^\sigma \,,
\end{equation}
where $\varepsilon_{\mu \nu \rho \sigma}$ is the totally antisymmetric Levi-Civita
tensor. The importance of the $\alpha_i$ arises as they vanish in one-loop-like
infra-red regions. Let us consider the dangerous infra-red regions associated to
the $\ell_1$ loop. The soft regions are given by
\begin{equation}
  \ell_1 \rightarrow p_1, \qquad \ell_1 \rightarrow p_1 + p_2,
\end{equation}
while the collinear regions are given by
\begin{equation}
  \ell_1 \rightarrow x p_1, \qquad \ell_1 \rightarrow p_1 + x p_2, \qquad \ell_1 \rightarrow p_1 + p_2 + x p_3.
\end{equation}
In each region, the loop momentum becomes a linear combination of $p_1, p_2$ and
$p_3$. Therefore, in each region, $\alpha_1$ vanishes. Similar argumentation
leads us to conclude that $\alpha_2$ vanishes in the infra-red regions
associated to $\ell_2$. Therefore, we see that the $\alpha_i$ are useful
functions to include in numerators in order to control infra-red behavior.

The next ingredient in our construction of evanescent integrals are numerator
functions which vanish for four-dimensional loop momenta. These are given by
\begin{equation}
  \mu_{ij} = \ell_i^{\epsilon} \cdot \ell_j^{\epsilon} = \frac{G\left(\begin{array}{ccccc}\!\ell_i\! &\!p_1\! &\!p_2\! &\!p_3\! &\!p_4\! \\ \!\ell_j\! &\!p_1\! &\!p_2\! &\!p_3\! &\!p_4\!\end{array}\right)}{2 G(1234)}.
\end{equation}
where $\ell_i^{\epsilon}$ are the $(D-4)$-dimensional piece of loop-momentum
$i$, transverse to the four-dimensional plane. Note that, by similar
argumentation as above, the $\mu_{ij}$ also vanish in infra-red regions and
therefore can also be used to control the infra-red behavior of the numerator.

Let us now discuss how we systematically construct a basis of evanescent
numerators that are independent modulo lower topologies. To begin, we consider
monomials of the $\alpha_i$ and $\mu_{jk}$, of the form
\begin{equation}
      m_\beta = \alpha_1^{\beta_1} \alpha_2^{\beta_2} \mu_{11}^{\beta_3} \mu_{22}^{\beta_4} \mu_{12}^{\beta_5}.
\end{equation}
We then construct the full set of monomials that satisfy renormalizable power
counting, vanishing in all infra-red regions and at least one factor of some
$\mu_{ij}$, thus resulting in an evanescent integral.
Altogether, we find 165 such monomials.

Finally, we impose the irreducibility constraint modulo lower
topologies. To do this, we work on the maximal cut in a numerical fashion. We
randomly sample the full set of $m_\beta$ over loop momenta which set the
propagators in Fig.~\ref{fig:dpTopology} to zero. Linear-dependencies of the
monomials on the maximum cut can be determined from linear dependencies of the
set of evaluations via linear algebra. Altogether, we find that a set of
numerators that is evanescent and independent on the maximal cut is
\begin{align}
  \begin{split}
  \!\!\!\bigg\{&\mu_{12}, \alpha_{2} \mu_{12}, \alpha_{2} \mu_{11}, 
  \alpha_{1} \mu_{12}, \alpha_{1} \mu_{22}, 
  \mu_{12}^2, \mu_{12} \mu_{22}, \mu_{11} \mu_{12}, 
  \mu_{11} \mu_{22},\alpha_{2}^2 \mu_{11}, \alpha_{1} \alpha_{2} \mu_{22},
  \\
  &\alpha_{1} \alpha_{2} \mu_{11},
  \alpha_{1}^2 \mu_{22}, \alpha_{2} \mu_{12}^2, 
  \alpha_{2} \mu_{12} \mu_{22}, \alpha_{2} \mu_{11} \mu_{12}, 
  \alpha_{2} \mu_{11}^2, \alpha_{1} \mu_{22}^2,
  \alpha_{1} \mu_{11} \mu_{12}, \mu_{12}^3, \\
  &
  \mu_{12}^2 \mu_{22}, 
  \mu_{12} \mu_{22}^2,\mu_{11} \mu_{12}^2, 
  \mu_{11}^2 \mu_{12}, \, \alpha_{1} \alpha_{2} \mu_{22}^2, 
  \, \alpha_{1} \alpha_{2} \mu_{11}^2, \, \alpha_{2} \mu_{12}^3,
  \, \alpha_{2} \mu_{12}^2 \mu_{22}, \\
  &\alpha_{2} \mu_{11} \mu_{12}^2, 
  \alpha_{2} \mu_{11}^2 \mu_{12}, \, \mu_{12}^4,
  \, \mu_{12}^3 \mu_{22}, \, \mu_{11} \mu_{12}^3\bigg\}\,.
\label{eq:DoublePentagonEvanescentBasis}
  \end{split}
\end{align}
Importantly, it is possible to show that these numerators actually span 
power-counting space. Note that monomials of the $\alpha_i$ which satisfy
renormalizable power counting are linearly independent and span a basis. It is
easy to count that there are 33 such monomials. As there are also 33 functions
in \eqref{eq:DoublePentagonEvanescentBasis}, we therefore find that the
monomials in \eqref{eq:DoublePentagonEvanescentBasis} can be used to express any
double-pentagon numerator which satisfies renormalizable power counting.

In summary, if the integrand basis in \eqref{eq:DoublePentagonEvanescentBasis}
is used for the double-pentagon topology, then all the associated integrals
contribute at $\mathcal{O}(\epsilon)$. It is therefore unnecessary to compute
the double-pentagon topology in computations of the finite parts of gauge theory
amplitudes.

\paragraph{Hexa-box Topology}
In the case of the hexa-box topology, similar logic to the above can be used.
Nevertheless, a direct application of the above argumentation leaves a
single element of the integrand basis which does not decouple in the $D
\rightarrow 4$ limit.
Therefore, we find that it is necessary to work at the level of master
integrals (and not at the integrand level).
Specifically, we search for a choice of the master integral which is
$\mathcal{O}(\epsilon)$, uniformly transcendental and of maximal weight.
If we can find such an integral, as it is maximal weight, its
coefficient in the amplitude cannot contain a factor of
$1/\epsilon$. Therefore we do not need to compute the
$\mathcal{O}(\epsilon)$ part of the integral and thus we can discard
it.
This search is greatly aided by the fact that there is only a single master
integral associated to the hexa-box topology, see Table \ref{tab:SixPointMasterIntegralCount}.
One such choice can be taken to be the hexa-box integral with the numerator
\begin{equation}
  \mathcal{N}_{\text{hb}}^{\epsilon, \text{UT}} =
  \epsilon^4 \mu_{11}
  G\left(\begin{array}{ccc}\!\ell_2\! &\!p_5\! &\!p_6\!  \\ \! p_1 \! &\! p_5 \! &\! p_6 \! \end{array}\right).
  \label{eq:decouplingHexabox}
\end{equation}
This numerator is a product of $\mu_{11}$ and a judiciously chosen Gram
determinant. Specifically, the Gram determinant vanishes in all infra-red
regions associated to the $\ell_2$ loop. Moreover, the numerator has
sufficiently low power-counting such that we do not encounter a double pole at
infinity. Therefore, we expect this integral to be both $\mathcal{O}(\epsilon)$,
and of uniform and maximal transcendental weight.
In order to validate that the integral is indeed of uniform and maximal
transcendental weight, we use \textsc{Kira} to numerically check that it can be written
as a linear combination of pure integrals, with coefficients that are
independent of the dimensional regulator.
This proves that the corresponding hexa-box topology can be ignored
when working with four dimensional external kinematics. Nevertheless,
since the inclusion of this integral topology in our computation does
not entail additional technical challenges, we keep it in the
following analysis as it allows us to embed all the remaining integral
topologies into it.
We refer to section~\ref{sec:CanonicalDEAndBasis} for discussion of a
pure basis for the hexa-box topology.

\section{Canonical Differential Equation and Pure Basis}
\label{sec:CanonicalDEAndBasis}

Let $J$ be a vector of
Feynman integrals associated with a given topology. It is well known
that the integrals satisfy a coupled first order differential equation
\begin{equation}
  \mathrm{d} J = {\bf \overline{M}}\, J,
  \label{eq:GenericDifferentialEquation}
\end{equation}
where ${\bf \overline{M}}$ is a matrix of differential one-forms that depend
rationally on the kinematics and the dimensional regulator $\epsilon = (4-D)/2$.
The construction of this system of differential equations
is a non-trivial task due both to the complexity of
the algebra required to obtain ${\bf \overline{M}}$ as well as to the
complexity of the form of ${\bf \overline{M}}$ itself. In order to
avoid simplify these two problems, it is useful to change to a basis $I$ of
so-called ``pure''~\cite{Henn:2013pwa} Feynman integrals, 
which satisfies a differential equation of the form
\begin{equation}\label{eq:CanonicalDifferentialEquation}
  \mathrm{d} I = \epsilon\, {\bf M} \,I \,,\qquad {\bf M} =  
  M_{\alpha} \,\mathrm{d}\log{W_\alpha}.
\end{equation}
The $M_{\alpha}$ are matrices of rational numbers and the $W_\alpha$
are algebraic functions of the kinematic invariants, often called letters. The
identification of such a pure basis reduces the computation of the
differential equation to the identification of the $W_\alpha$,
alongside the determination of the set of rational-number matrices
$M_{\alpha}$. Both of these components can then be determined via
finite-field reconstruction
methods~(see e.g.~Refs.~\cite{Abreu:2018rcw,Abreu:2020jxa,Abreu:2021smk}
for a detailed discussion in the context of five-point integrals), drastically
simplifying the computation of the differential equation.

Let $I$ be related to $J$ through
$I = U^{-1} J$. The change-of-basis matrix $U$ is invertible and
rational in $\epsilon$, but potentially algebraic in the kinematic
invariants. The matrices ${\bf M}$ and ${\bf \overline{M}}$ are
related by the following change-of-basis relation
\begin{equation}\label{eq:ChangeOfBasis}
  {\bf M} = U^{-1}{\bf
  \overline{M}}  U - U^{-1} \dd U\,.
\end{equation}

In this section, we discuss the details of the construction of the
differential equation and the pure basis for the set of topologies
embedded in the hexa-box topology defined in
section~\ref{sec:IntegralDefinitions}, see Fig.~\ref{fig:hboxParam}.
In particular, we focus on novel difficulties in construction the
differential equation itself due to the four-dimensional nature of the
external kinematics.
The hexa-box topology contains a total of 202 master integrals. 
While the hexa-box integral itself is not necessarily needed
for four-dimensional gauge theories, as argued in the previous section,
we nevertheless include it also in our basis since it is trivial
to find a pure representative and this will allow us to explicitly test
the argument given in section \ref{sec:9PropagatorIrrelevance}
for the decoupling of this nine-propagator integral.

\subsection{Approach to Canonicalization}

In order to determine a pure basis of six-point massless Feynman integrals, we
follow the general two-step strategy of first constructing a basis of
integrals that is nearly pure, and then using the structure of the differential
equation to construct a change of basis matrix to a pure basis.
This strategy is not new, and we summarise it here mainly to establish our 
notation.

We begin by identifying the integrals that already appear in
the five-point one-mass case. For those, we take the pure masters
integrals identified in \cite{Abreu:2020jxa}. 
For the genuine
six-point topologies, it is simple to identify candidate
pure master integrals for the hexa-box, penta-triangle, hexa-bubble 
and penta-box topologies, following ideas similar to those used in 
e.g.~\cite{Abreu:2018rcw,Abreu:2018aqd,Abreu:2020jxa}. Explicitly,
we take the following candidate master integrals for these topologies:
\begin{itemize}
  \item For the hexa-box topology:
\begin{equation}\label{eq:pureCandidate_hb}
\mathcal{N}_1^{(\text{hb})} = \epsilon^4\mu_{11}[1234]\sqrt{\Delta^{(5)}_{1234}}(\ell_1 + p_6)^2 s_{56}\,.\\
\end{equation}
  \item For the penta-triangle topology:
\begin{equation}\label{eq:pureCandidate_pt}
\mathcal{N}_1^{(\text{pt})} = \epsilon^4 \mu_{11}[2346]
\sqrt{\Delta^{(5)}_{2346}}\,.
\end{equation}
  \item For the penta-box topology:
\begin{align}\begin{split}\label{eq:pureCandidate_pb}
  \mathcal{N}_1^{(\text{pb})} &= 
  \epsilon^4 s_{12} s_{23} s_{56} (\ell_1 + p_6)^2\,, \\
  \mathcal{N}_2^{(\text{pb})} &= 
  \epsilon^4  s_{56}\mu_{11}[3216] \sqrt{\Delta^{(5)}_{3216}}\,, \\
  {\mathcal{N}}_3^{(\text{pb})} &= -
  \epsilon^4 \left(\frac{1-2\epsilon}{1+2\epsilon} \right)
  \left(\mu_{11}[3216]\mu_{22}[3216] - \mu_{12}^2[3216]\right) 
  \sqrt{\Delta^{(5)}_{3216}}\,.
\end{split}\end{align}
  \item For the hexa-bubble topology:
\begin{equation}\label{eq:pureCandidate_hbub}
  \mathcal{N}_1^{(\text{hbub})} = \epsilon^3 \mu_{11}[1234]
  \sqrt{\Delta^{(5)}_{1234}} \frac{(\ell_1+p_6)^2}{(\ell_1+\ell_2)^2}.
\end{equation}
\end{itemize}
In these expressions, we have used 
\begin{equation}
  \mu_{ij}[k_1k_2k_3k_4]=
   \frac{G\left(\begin{array}{ccccc}\!\ell_i\! &\!p_{k_1}\! &\!p_{k_2}\! &\!p_{k_3}\! &\!p_{k_4}\! \\ \!\ell_j\! &\!p_{k_1}\! &\!p_{k_2}\! &\!p_{k_3}\! &\!p_{k_4}\!\end{array}\right)}
  {2 G(k_1k_2k_3k_4)}.
\end{equation}
We remark that in the case that the external momenta are four-dimensional, the
$\mu_{ij}[k_1 k_2 k_3 k_4]$ do not depend on the choice of $k_1,k_2,k_3,k_4$.
While the explicit expressions might look different, they are equivalent
modulo the constraints of Eq.~\eqref{eq:ISPDependenceConstraint} and
$\Delta_6=0$.

Given the number of masters integrals, the double-box topology
of Fig.~\ref{fig:dbTopology} is the most complicated topology we need 
to address. Some pure integrals are easy to identify by standard
techniques, and we complete this set with masters related to the on-shell pure
basis of \cite{Henn:2021cyv}. 
Explicitly, we take as candidate master integrals the following insertions
\begin{align}
  \begin{split}\label{eq:pureCandidate_db}
  \mathcal{N}^{(\text{db})}_1 &= \epsilon^4 s_{23} s_{345} s_{56}\,, \\
  \mathcal{N}^{(\text{db})}_2 &= \epsilon^4 s_{23} s_{56} (\ell_1+p_6)^2\,, \\
%odd[N4]
  \mathcal{N}^{(\text{db})}_3 &= \epsilon^4
\frac{s_{56}}{s_{25} s_{35}} \mathrm{tr}_5(p_3, \ell_1 - p_{12}, \ell_1 - p_{12} - p_2, p_5, p_3, p_5, p_2, p_{16})\,, \\
% odd[N6]
  \mathcal{N}^{(\text{db})}_4 &= \epsilon^4
\frac{s_{23}}{s_{35} s_{36}} \mathrm{tr}_5(p_5, \ell_2 - p_6, \ell_2 - 2 p_6, p_3, p_5, p_3, p_6, p_{12})\,, \\
% even[N8]
  \mathcal{N}^{(\text{db})}_5 &= \epsilon^4
\frac{1}{s_{25} s_{35} s_{36}}\mathrm{tr}_5(p_3, \ell_1 \!-\! p_{12}, \ell_1 \!-\! p_{12} \!-\! p_2, p_5, p_3, \ell_2 \!-\! 2 p_6, \ell_2 \!-\! p_6, p_5, p_3, p_5, p_2, p_6) , \\
  \mathcal{N}^{(\text{db})}_6 &= \epsilon^3(1-2\epsilon)
  \mu_{11}[1234]\,, \\
  \mathcal{N}^{(\text{db})}_7 &= \epsilon^4 \mu_{12}[2356] \Delta_{2356}^{(5)}\,.
  \end{split}
\end{align}
Here we used the $\gamma_5$ prescription of Ref.~\cite{Korner:1991sx},
so that
\begin{equation}
  \mathrm{tr}_5(k_1 \cdots k_n)
  = \frac{1}{96}\mathrm{tr}(\slashed{k_1} \cdots \slashed{k_n} \gamma^{\mu_{n+1}} \gamma^{\mu_{n+2}} \gamma^{\mu_{n+3}} \gamma^{\mu_{n+4}} ) \varepsilon_{\mu_{n+1} \mu_{n+2} \mu_{n+3} \mu_{n+4}}\,.
\end{equation}
To convert this into a Lorentz-invariant expression, we perform the trace of the
chain of Dirac matrices, and make the replacement $\epsilon(p_i p_j p_k p_l)
\rightarrow \sqrt{\Delta^{(5)}_{ijkl}}$.

With the choice of basis described above, the differential
equation~\eqref{eq:GenericDifferentialEquation} takes the following
form
\begin{equation}\label{eq:EpsLinearDifferentialEquation}
\dd J = \left({\bf \overline{M}}^{(0)} + \epsilon {\bf
    \overline{M}}^{(1)}\right) \,J\,.
\end{equation}
That is, the entries of $J$ are in general not pure, but satisfy 
differential equations
that are linear in $\epsilon$. This is already a good starting point, and
we now discuss how to construct a change of basis $I =
U^{-1} J$ such that $I$ satisfies a canonical differential equation of
the form~\eqref{eq:CanonicalDifferentialEquation}.
In the new basis, the differential equation matrix ${\bf M}$ is
related to the one in Eq.~\eqref{eq:EpsLinearDifferentialEquation} by
Eq.~\eqref{eq:ChangeOfBasis}, namely
\begin{equation}
 {\bf M} = \epsilon U^{-1}{\bf
  \overline{M}}^{(1)}  U+ U^{-1}\left({\bf
  \overline{M}}^{(0)}  U- \dd U\right)\,.
\end{equation}
In other words, for the system to be in canonical form, $U$ should
satisfy the equation
\begin{equation}\label{eq:dU}
\dd U = {\bf \overline{M}}^{(0)}  U\,.
\end{equation}
We observe that the ${\cal O}(\epsilon^0)$ term in
Eq.~\eqref{eq:EpsLinearDifferentialEquation} is a lower-triangular
matrix.
We then find it convenient to write it as
\begin{equation}\label{eq:M0}
{\bf \overline{M}}^{(0)} = {\bf \overline{M}}_{D}^{(0)} + {\bf \overline{M}}_{L}^{(0)} \,,
\end{equation}
where the matrix ${\bf \overline{M}}_{D}^{(0)}$ is diagonal while
${\bf \overline{M}}_{L}^{(0)}$ is strictly lower triangular, and to
parametrise $U$ as the product
\begin{equation}
U = U_D\,U_L\,.
\end{equation}
It is easy to show that a solution to Eq.~\eqref{eq:dU} is given by
\begin{align}
  \dd U_D &= {\bf \overline{M}}_{D}^{(0)}  U_D\,, \label{eq:dUD}\\
  \dd U_L &= \left(U_D^{-1}{\bf \overline{M}}_{L}^{(0)} U_D \right)
            U_L \,. \label{eq:dUL}
\end{align}
The above equations clarify the role of the two matrices $U_D$ and
$U_L$. The former is a diagonal matrix that fixes the normalisation of
the integrals ensuring that the diagonal entries of ${\bf M}$ are
$\epsilon$ factorised. The latter fixes the non-diagonal entries and
is given by the sum of the identity matrix and a strictly
lower-triangular matrix.

The above decomposition allows us to proceed in two steps. We first
determine $U_D$ by solving  Eq.~\eqref{eq:dUD}, and then we determine
$U_L$ by solving Eq.~\eqref{eq:dUL}.
Given that ${\bf \overline{M}}_{D}^{(0)}$ is diagonal, a particular
solution to Eq.~\eqref{eq:dUD} is
\begin{equation}\label{eq:UD-solution}
U_D = \Lambda_D \exp\left(\int {\bf \overline{M}}_{D}^{(0)}\right)\,,
\end{equation}
where $\Lambda_D$ is a constant matrix and the integral is defined such
that
\begin{equation}\label{eq:integral-definition}
\dd \int {\bf \overline{M}}_{D}^{(0)} = {\bf \overline{M}}_{D}^{(0)}\,,
\end{equation}
and we set all corresponding integration constants to zero. Since
${\bf \overline{M}}_{D}^{(0)}$ is diagonal, the exponential in
Eq.~\eqref{eq:UD-solution} amounts to the diagonal matrix whose
entries are the exponentials of the entries of
${\bf \overline{M}}_{D}^{(0)}$.
The choice of $\Lambda_D$ is arbitrary, and we choose it to be the identity
matrix so that $U_D$ simply acts as a normalisation on the integrals
and does not rotate the basis. We will discuss how to evaluate the
integral in Eq.~\eqref{eq:UD-solution} later on.

We now switch to Eq.~\eqref{eq:dUL}. Given that now the matrix
${\bf \overline{M}}_{L}^{(0)} $ is not diagonal, a particular solution
to Eq.~\eqref{eq:dUL} can be obtained by means of a Magnus
expansion~(see e.g.~\cite{Argeri:2014qva}). By inspection of
${\bf \overline{M}}_{L}^{(0)} $, we observe that the corresponding
Magnus exponential does not truncate at first order. 
To simplify the procedure, we can further express the matrix
$U_D^{-1}\overline{\bf M}_{L}^{(0)}U_D$ as
\begin{equation}
\tilde{\bf M}_{L}^{(0)} \equiv U_D^{-1}\overline{\bf M}_{L}^{(0)}U_D=
\tilde{\bf M}_{L}^{(0),\,{\rm on}}+\tilde{\bf M}_{L}^{(0),\,{\rm off}}\,,
\end{equation}
where the two matrices in the r.h.s.~are derived from the
corresponding on-shell and
off-shell differential
equations, $\tilde{\bf M}_{L}^{(0),\,{\rm on}}$ and
$\tilde{\bf M}_{L}^{(0),\,{\rm off}}$ respectively.
Accordingly, we can express the matrix $U_L$ as
\begin{equation}\label{eq:UL-deco}
U_L = U_L^{\rm on}\,U_L^{\rm off}\,.
\end{equation}
The differential equation~\eqref{eq:dUL} can then be solve by
requiring that
\begin{align}
  \dd U_L^{\rm on} &=\tilde{\bf M}_{L}^{(0),\,{\rm on}} U_L^{\rm on}\,, \label{eq:dULon}\\
  \dd U_L^{\rm off} &= \left(\left(U_L^{\rm on}\right)^{-1}\tilde{\bf
                      M}_{L}^{(0),\,{\rm off}} U_L^{\rm on} \right)
            U_L^{\rm off} \,. \label{eq:dULoff}
\end{align}
The main advantage of the decomposition in Eq.~\eqref{eq:UL-deco} is
that the Magnus expansion that solves each of
Eqs.~\eqref{eq:dULon} and ~\eqref{eq:dULoff} now truncates at first
order. That is,
\begin{align}
U_L^{\rm on} &= \Lambda_L^{\rm on} \exp\left(\int {\bf
    \tilde{M}}_{L}^{(0),\,{\rm
  on}}\right)\,, \label{eq:ULon-solution}\\
U_L^{\rm off} &= \Lambda_L^{\rm off} \exp\left(\int \left(U_L^{\rm on}\right)^{-1}\tilde{\bf
                      M}_{L}^{(0),\,{\rm off}} U_L^{\rm on}\right)\,, \label{eq:ULoff-solution}  
\end{align}
where the integration in the exponent is defined analogously to
Eq.~\eqref{eq:integral-definition}.
As before, without loss of generality, we set the matrices of
integration constants $\Lambda_L^{\rm on}$ and $\Lambda_L^{\rm off}$
to the identity matrix.

This procedure simplifies the computation of the change-of-basis
matrix $U$, in that we can proceed in stages by computing each of
the factors $U_D$, $U_L^{\rm on}$ and $U_L^{\rm off}$. To this end, we pursue an
analytic reconstruction approach to this determination, that we describe in the
next section.
We apply this procedure on a sector-by-sector
basis, for each of the six-point sectors.

\subsection{Transformation to canonical basis by analytic reconstruction}

In order to use
Eqs.~\eqref{eq:UD-solution},~\eqref{eq:ULon-solution} and~\eqref{eq:ULoff-solution},
we need to determine the matrices
${\bf \overline{M}}_{D}^{(0)}$,
$ {\bf \tilde{M}}_{L}^{(0),\,{\rm on}}$ and
$ \left(U_L^{\rm on}\right)^{-1}\tilde{\bf M}_{L}^{(0),\,{\rm off}}
U_L^{\rm on}$. We will do this via analytic reconstruction.
For simplicity, we shall denote any of these matrices by
$\mathfrak{m}$. 
To achieve this, we will exploit the expectation that the matrices
$U_D, U_L^{\text{on}}$ and $U_L^{\text{off}}$ have algebraic entries. This puts
strong constraints on 
the analytic structure of $\mathfrak{m}$. In this section, we will use these
constraints to construct compact ansaetze for the matrices $\mathfrak{m}$,
leading to a highly efficient analytic reconstruction procedure for the change
of basis matrix $U$ that will allow us to obtain a pure basis. 

\paragraph{Singularity Structure of the Pre-Canonical Differential Equation}
We begin by determining the singularity structure of the entries of
the matrices $\mathfrak{m}$.
By construction, the entries of $\mathfrak{m}$ are algebraic functions of the
external kinematics.
Similar to the strategy commonly used for master integral
coefficients in amplitudes (see e.g.~\cite{Abreu:2018zmy}), we make an
ansatz that these functions can be written in a form where the denominator is
given by a product of even letters.\footnote{Roughly speaking, even letters
are those that do not involve a square root. We will return to this topic below,
but we stress that this is not a
restriction as the odd letters themselves (those involving roots) 
can be rewritten in a form with
algebraic numerators and a denominator that is a product of even letters.} That
is,
\begin{equation}
  \mathfrak{m}_{ij} = \frac{\mathfrak{n}_{ij}}{\prod_{k \in \text{even}} W_k^{q_{ijk}}},
  \label{eq:DEStructureAnsatz}
\end{equation}
where $\mathfrak{n}_{ij}$ is a polynomial one form that potentially also has
overall factors of the square roots. These square root factors arise from the
form of the basis $J$ in \eqref{eq:EpsLinearDifferentialEquation} and can be
read off by inspection.
We note that this ansatz places non-trivial constraints on the analytic
structure of the $\mathfrak{m}_{ij}$, but nevertheless we find it to be true
via numerical experimentation.
We construct the set of even letters that we use in \eqref{eq:DEStructureAnsatz}
in two steps. Firstly, we take the set of planar five-point one-mass letters
from Ref.~\cite{Abreu:2020jxa}. We rewrite them in six-particle kinematics,
maintaining planarity and considering all possible identifications of the
off-shell leg with a consecutive pair of massless legs.
We then complete this set using the even letters found on the maximal cut in
Ref.~\cite{Henn:2021cyv}.
Importantly, we enforce that all $W_k$ are (parity-invariant) polynomials in
Mandelstam variables. In a number of cases, even letters of
Ref.~\cite{Abreu:2020jxa} have been chosen to be square roots of polynomials in
Mandelstams and so we simply choose to take our $W_k$ as the polynomial under
the square root. A more delicate subtlety arises when considering even letters 
of Ref.~\cite{Henn:2021cyv}, which makes use of chiral objects such as
spinor brackets or the Levi-Civita tensor. We therefore need to translate the
expressions of Ref.~\cite{Henn:2021cyv} to our conventions. In two cases, this
requires non-trivial analysis. The first is
\begin{equation}
  \tilde{W}_{145} = \langle 12 \rangle [23] \langle 34 \rangle [45] \langle 56 \rangle [61]
- [12] \langle 23 \rangle [34] \langle 45 \rangle [56] \langle 61 \rangle,
\end{equation}
which is the leading singularity of the massless one-loop hexagon integral.
Notably, $\tilde{W}_{145}$ is written in terms of spinor-helicity variables (we
refer to Ref.~\cite{Henn:2021cyv} for their definitions). As $\tilde{W}_{145}$
is chiral, it is not expressible in terms of Mandelstam variables. To make use
of this letter, we simply work with its square which is parity invariant. That
is, we define
\begin{equation}
    W_{145} = (s_{12} s_{234} s_{45} + s_{23} s_{345} s_{56} + s_{34} s_{123} s_{16} - s_{123} s_{234} s_{345})^2 - 4 s_{12} s_{23} s_{34} s_{45} s_{56} s_{16}.
    \label{eq:W145SquareDef}
\end{equation}
which satisfies $W_{145} = \tilde{W}_{145}^2$.
The second case requiring non-trivial analysis is the collection of letters
given by Levi-Civita symbols. Once again, we rewrite them in terms of expressions
that only depend on Mandelstam invariants by working with their square. That is,
we make use of the identity
\begin{equation}
  \mathrm{d}\log[\epsilon(ijkl)] = \frac{1}{2} \mathrm{d}\log[G(ijkl)].
\end{equation}
In this way, we trade parity odd letters from Ref.~\cite{Henn:2021cyv}
for equivalent expressions at the level of $\mathrm{d}\log$s.

In order to determine the exponents $q_{ijk}$, we use a univariate-slice
strategy~\cite{Abreu:2018zmy}. To this end, we must construct a univariate slice
in the space of the Mandelstam invariants, parametrised by a variable $t$.
In principle, this is a non-trivial problem, as the slice of phase-space must be
four-dimensional, i.e.~every point on the slice must satisfy
Eq.~\eqref{eq:SixPointGramConstraint}. In order to solve this problem, we make
use of the univariate-slice generation strategy of Ref.~\cite{Abreu:2021asb}.
Here, a set of Mandelstams of the form
\begin{equation}
  \vec{s}(t) = \vec{a} + \vec{b} t,
  \label{eq:SliceParameterization}
\end{equation}
is generated by starting from a multi-line BCFW shift~\cite{Elvang:2008vz} in
spinor space. As spinors are inherently four-dimensional, this approach
results in numerical values of $\vec{a}$ and $\vec{b}$ such that $\vec{s}(t)$
satisfies Eq.~\eqref{eq:SixPointGramConstraint} for all values of $t$.

Given the information on the denominators of the letters, we now combine it with
constraints that arise from the structure of the change of basis matrices.
Specifically, we assume that $U_D$, $U_L^{\text{on}}$ and $U_L^{\text{off}}$ are
all algebraic functions of the Mandelstam variables.
Given the discussion in the previous section, they are all expressed as
exponentials of integrals of the matrices which we wish to reconstruct.
It is therefore important to consider how one computes the exponential of a
matrix. In practice, we have two distinct cases: either the exponent matrix is
diagonal, or strictly lower triangular. We will consider each case in turn.

\paragraph{Ansaetze for The Diagonal Case}
We consider first the diagonal case. Here, the matrix exponential is calculated
by taking the exponential of the entries along the diagonal. This implies that
the entries of $\overline{\bf M}_D^{(0)}$
must be given by derivatives of
logarithms. Given that we have validated and determined the denominator structure of
Eq.~\eqref{eq:DEStructureAnsatz} using a univariate slice, we therefore use an
ansatz of the form
\begin{equation}
  \mathfrak{m}_{ij} = \sum_{k \in \text{even}} c_{ijk} \mathrm{d} \log(W_k).
  \label{eq:dlogAnsatz}
\end{equation}
In practice, only a single entry is non-zero and is easily reconstructed with
this strategy. 
Importantly, by using such an ansatz, we bypass the much more complicated
numerator reconstruction of the ansatz in Eq.~\eqref{eq:DEStructureAnsatz}.
Moreover, the $\mathrm{d}\log$ ansatz of Eq.~\eqref{eq:dlogAnsatz} greatly
facilitates the computation of the integral in Eq.~\eqref{eq:UD-solution}, as it is
explicitly written in the form of a total derivative. Nevertheless a subtlety
arises when computing the integral, as we find that some $c_{ijk}$ associated to
$\mathrm{d}\log$s of Gram determinants and $W_{145}$ are half integer.
In order to avoid introducing square-root branch cuts associated to these
letters into our basis definitions, we rewrite the $\mathrm{d}\log$ expression
by making use of the identities
\begin{align}
  \frac{1}{2}\mathrm{d}\log\left[  G\left( a b c d \right)\right]
  &= \mathrm{d}\log\left[ G\left(
\begin{array}{cccc}
  \!a\!&\!b\!&\!c\!&\!d\! \\
  \!1\!&\!2\!&\!3\!&\!4\!
\end{array}
  \right) \right]
    - \frac{1}{2}\mathrm{d}\log\left[  G\left( 1234 \right) \right]\,,
  \\
  \frac{1}{2}\mathrm{d}\log[ W_{145} ]
  &= \mathrm{d}\log\left[ W_{145}^{ijkl} \right]
     - \frac{1}{2}\mathrm{d}\log\left[ G(i j k l) \right]\,,
\end{align}
where we define
\begin{equation}
    W_{145}^{ijkl} = \epsilon(p_i p_j p_k p_l) \tilde{W}_{145},
    \label{eq:MultiW145Definition}
\end{equation}
which is again a polynomial in Mandelstam variables as it is explicitly
parity invariant. Using these identities, we are able to rewrite the entries of
$\overline{\bf M}_D^{(0)}$ such that the only half integer in the
$\mathrm{d}\log$ representation is associated to $\mathrm{d}\log[G(1234)]$, and
therefore no extra branch cuts are introduced.

\paragraph{Ansaetze for the Lower Triangular Case}
Let us now consider the strictly lower triangular case. Here, the matrix
exponential can be calculated using the power series form, as strictly
lower-triangular matrices are nilpotent and the series truncates. We therefore
conclude that the argument of the exponential is an algebraic function, and so
our target matrices must be derivatives of algebraic functions.
That is, we will write that
\begin{equation}
  \mathfrak{m}_{ij} = \mathrm{d} \left[\frac{\tilde{\mathfrak{n}}_{ij}}{\prod_{k \in \text{even}} W_k^{\tilde{q}_{ijk}}}\right],
  \label{eq:TotalDerivativeAnsatz}
\end{equation}
where the $\tilde{q}_{ijk}$ are integers, and $\tilde{\mathfrak{n}}_{ij}$ is a
polynomial function of both the Mandelstams and the roots in the problem.
In this context, we must again consider how to treat $W_{145}$. An important
subtlety arises from the fact that $W_{145} = \tilde{W}_{145}^2$ and hence, when
$\tilde{W}_{145}$ vanishes, $W_{145}$ will vanish quadratically. Therefore, an
ansatz which makes use of the expression $W_{145}$ in
Eq.~\eqref{eq:W145SquareDef}, does not allow for the possibility of the argument
blowing up linearly when $W_{145} = 0$. To rectify this, we instead use
$W_{145}^{1234}$ in place of $W_{145}$ in Eq.~\eqref{eq:TotalDerivativeAnsatz},
a fact that we will suppress for notational convenience.
A further subtlety in using Eq.~\eqref{eq:TotalDerivativeAnsatz} as an ansatz is
that the argument of the derivative can be shifted by a constant function, and
not change $\mathfrak{m}_{ij}$. However, such a constant can be absorbed into
the integration constant matrices $\Lambda_L^{\text{on}}$ and
$\Lambda_L^{\text{off}}$, so we are free to set this degree of freedom in the
ansatz to zero.

In order to determine the $\tilde{q}_{ijk}$ we will again make use of a
univariate-slice procedure.
As we are working with algebraic functions satisfying constraints,
it turns out that the action of the total derivative can introduce denominator
factors that are not simply higher powers of the $W_k$, and so
determining the $\tilde{q}_{ijk}$ requires a careful analysis.
A simple first observation is that, if $\tilde{\mathfrak{n}}$ depends on a
square-root, then the action of the derivative naturally introduces a factor of
this root as, for any function $\Delta$
\begin{equation}
    \mathrm{d} \sqrt{\Delta} = \frac{\mathrm{d} \Delta}{2 \sqrt{\Delta}}\,.
\end{equation}
We therefore see that, if the $\tilde{\mathfrak{n}}_{ij}$ depend on a root, upon
acting with the total derivative, we will find this root in the denominator.
Next, we consider a consequence of working under the constraint $\Delta_6 = 0$.
Note that, as we work with this constraint we also have that $\mathrm{d}\Delta_6
= 0$. We therefore have a constraint on the differentials with respect to the 9
Mandelstam invariants. By writing these in terms of an independent set of 8, we
introduce an extra denominator factor. Specifically, we use this constraint to
eliminate $\mathrm{d} s_{16}$, finding
\begin{equation}
      \mathrm{d} s_{16} = - \frac{1}{\frac{\partial \Delta_6}{\partial s_{16}}} \left[ \sum_{i=1}^{5} \left(\frac{\partial}{\partial {s_{i i+1}}} \Delta_6 \right) \mathrm{d} s_{i i+1} + \sum_{i=1}^{3} \left(\frac{\partial}{\partial {s_{i i+1 i+2}}} \Delta_6 \right) \mathrm{d} s_{i i+1 i+2} \right].
\end{equation}
We therefore see that by solving this equation, expressions for the total
derivative of a given function will involve a denominator factor
of
\begin{equation}
   \frac{\partial}{\partial s_{16}} \Delta_6 = 2 G\left( \begin{array}{cccc} \!1\!& \!2\!&\!3\!&\!4\! \\ \!3\!& \!4\!&\!5\!&\!6\! \end{array} \right).
\end{equation}
Indeed, this is not so surprising as we are essentially taking
$s_{16}$ to be a solution of a quadratic equation. This implicitly
makes use of a root and so we similarly end up with an expression in
the denominator.\footnote{It is perhaps more unexpected that the
  expression is again a Gram determinant. However, we note that all
  derivatives of $\Delta_6$ with respect to planar Mandelstam
  variables are proportional to two-line Gram determinants.}

A much more subtle complication lies in the interaction of the previous two
points. If we consider the case where the square root is $\sqrt{\Delta_{5}}$,
then it turns out that this is not the only algebraic object that
$\tilde{\mathfrak{n}}_{ij}$ can depend on which can generate these denominators.
Specifically, one can show that
\begin{equation}
  \mathrm{d} \sqrt{\Delta^{(5)}_{i_1 i_2 i_3 i_4}}
  = \frac{
        G\left( \begin{array}{cccc} \!1\! & \!2\! &\!3\! &\!4\! \\ \!3\! & \!4\! &\!5\! &\!6\! \end{array} \right)
        \mathrm{d} G\left( \begin{array}{cccc} \!i_1\! & \!i_2\! &\!i_3\! &\!i_4\! \\ \!1\! & \!2\! &\!3\! &\!4\! \end{array} \right)
        - \frac{1}{2} G\left( \begin{array}{cccc} \!i_1\! & \!i_2\! &\!i_3\! &\!i_4\! \\ \!3\! & \!4\! &\!5\! &\!6\! \end{array} \right) \mathrm{d} G\left( 1234 \right)
    }{\sqrt{\Delta_{5}}G\left( \begin{array}{cccc} \!1\! & \!2\! &\!3\! &\!4\! \\ \!3\! & \!4\! &\!5\! &\!6\! \end{array} \right)}.
\end{equation}
Here, we see that the numerator of this fraction is a polynomial one form, as the
first term explicitly contains a factor to cancel the denominator in the
derivative, while in the second term, the derivative acts on $G(1234)$, which is
independent of $s_{16}$.
For this reason, in cases where $\sqrt{\Delta}_5$ is involved, we are led to
consider an ansatz for the $\tilde{\mathfrak{n}}_{ij}$ which 
is polynomial not just in the
Mandelstams, but also in the independent variables $\sqrt{\Delta^{(5)}_{ijkl}}$.
A perhaps surprising conclusion is that such a ``polynomial'', given
Eq.~\eqref{eq:ExtraGramRootDefinition}, can be expressed as a rational function.
We expect this to be a general complication that
arises when working with such complex algebraic functions.

In summary, assuming that the differential equation entry $\mathfrak{m}_{ij}$
depends on two square roots $\sqrt{\Delta_1}$ and $\sqrt{\Delta_2}$, we see
that we can expand the derivative in Eq.~\eqref{eq:TotalDerivativeAnsatz} to read
\begin{equation}
\mathrm{d} \left[\frac{\tilde{\mathfrak{n}}_{ij}}{\prod_{k \in \text{even}} W_k^{\tilde{q}_{ijk}}}\right] = 
\frac{\overline{\mathfrak{n}}_{ij}}{G\left( \begin{array}{cccc} \! 1 \! &\! 2 \! &\! 3 \! &\! 4 \! \\ \! 3 \! &\! 4 \! &\! 5 \! &\! 6 \!\end{array} \right) \sqrt{\Delta_1} \sqrt{\Delta_2} \prod_{k \in \text{even}} W_k^{\tilde{q}_{ijk} + 1}}.
\end{equation}
Naturally, in cases where $\Delta_1$ or $\Delta_2$ are not present, their roots can be
omitted from the denominator. Given this result, the $\tilde{q}_{ijk}$ can
easily be determined from a univariate slice procedure.
Furthermore, we take an ansatz for the $\tilde{\mathfrak{n}}_{ij}$ 
which is polynomial
in the Mandelstam variables and the square roots in the problem. If the square
root is $\sqrt{\Delta_5}$, we also allow it to be polynomial in the independent
$\sqrt{\Delta_{i_1 i_2 i_3 i_4}^{(5)}}$.

A non-trivial step is constructing a linearly independent ansatz for a
polynomial when working under the six-point Gram constraint,
Eq.~\eqref{eq:SixPointGramConstraint}. It is clear that, for a polynomial of
degree $P$, the constraint will introduce linear relations between the
$\vec{s}^{\vec{\alpha}}$ monomials of dimension $P$.
However, it turns out that we can efficiently find a subset of these monomials
which are linearly independent.
We make use of a procedure introduced in Ref.~\cite[Section
2.2]{DeLaurentis:2022otd}, which we review here and adapt to the case of a
single constraint. We refer the reader to the aforementioned reference for
further details and proofs.
First, we introduce a monomial ordering $\succ$, such as the
degree-reverse-lexicographic ordering, on the set
of monomials in $\vec{s}$. This allows us to rewrite $\Delta_6$ as
\begin{equation}
  \Delta_6 = \mathrm{LT}(\Delta_6) + \tau(\Delta_6),
\end{equation}
where $\mathrm{LT}(\Delta_6)$ is the so-called ``lead term'' of $\Delta_6$, and
$\tau(\Delta_6)$ is a linear combination of monomials that are all lower in the monomial ordering.
For any polynomial $p$, we can now systematically apply the $\Delta_6 = 0$
relation by repeatedly replacing any instances of $\mathrm{LT}(\Delta_6)$ with
$-\tau(\Delta_6)$. This procedure eventually terminates due to properties of the
monomial ordering and results in a representation of $p$ which is a linear
combination of monomials, none of which has a factor of
$\mathrm{LT}(\Delta_6)$. In conclusion, a unique representation of 
any polynomial $p$ can be taken to be
\begin{equation}
  p(\vec{s}) = \sum_{\substack{\mathrm{LT}(\Delta_6) \, \nmid \, \vec{s}^{\vec{\alpha}}  \\ \sum_i \alpha_i = P}} c_{\vec{\alpha}} \vec{s}^{\vec{\alpha}},
  \label{eq:fourDPolynomialAnsatz}
\end{equation}
where the $c_{\vec{\alpha}}$ are unknown rational numbers and the sum is over
all monomials of degree $P$ which do not have a factor of $\mathrm{LT}(\Delta_6)$.
Let us comment that at higher multiplicities, there are multiple gram
constraints that one must consider. The above procedure is easily generalized to
such cases, where one must compute a Groebner basis associated to the Gram
constraints. We refer the reader to Ref.~\cite{DeLaurentis:2022otd} for more
details in the context of polynomials in spinor brackets.

\paragraph{Reconstruction and Summary}

Given this procedure to construct a non-redundant ansatz for our unknown
polynomials, it remains to determine the $c_{\vec{\alpha}}$. To this end, we
constrain them using finite-field evaluations of the differential equation.
In summary, our procedure for the reconstruction of the change of basis proceeds as follows:
\begin{enumerate}
  \item We determine the singularity structure of the $\mathfrak{m}$.
  \item We use this information to build two classes of ansaetze for the
        $\mathfrak{m}$, either $\mathrm{d}\log$ form ansaetze or total derivatives of
        algebraic functions. Due to the presence of the six-point Gram constraint,
        Eq.~\eqref{eq:SixPointGramConstraint}, this presents new theoretical subtleties.
  \item We constrain the ansatz with finite-field evaluations of the differential equation.
  \item We perform the integrals in Eqs.~\eqref{eq:UD-solution}, \eqref{eq:ULon-solution} and \eqref{eq:ULoff-solution}, by exploiting that
        our ansaetze are total derivatives and compute the exponential by using
        that the argument is a nilpotent matrix.
\end{enumerate}

In practice, we find that it is fruitful to build constraints by
sequentially gathering constraints from the differential equation
matrix with respect to each variable in $\vec{s}$, and that it is not
necessary to examine all variables.
We close by stressing that ansaetze of the form of Eq.~\eqref{eq:dlogAnsatz} or
Eq.~\eqref{eq:TotalDerivativeAnsatz} have orders of magnitude fewer unknowns
than that of Eq.~\eqref{eq:DEStructureAnsatz}, greatly decreasing the number of
numerical evaluations of the differential equation that one must perform.

%%%%%%%%%%%%%%%%%%%%%%%%%%%%%%%%%%%%%%%%%%%%%%%%%%%%%%%%%%%%%%%%%%%%%%%%%%%

%%%%%%%%%%%%%%%%%%%%%%%%%%%%%%%%%%%%%%%%%%%%%%%%%%%%%%%%%%%%%%%%%%%%%%%%%%%

%%%%%%%%%%%%%%%%%%%%%%%%%%%%%%%%%%%%%%%%%%%%%%%%%%%%%%%%%%%%%%%%%%%%%%%%%%%

\section{Analytic Differential Equation and Numerical Checks}

Using the strategy discussed in the previous section, we obtain
a differential equation in canonical form for the 
hexa-box topology of Fig.~\ref{fig:hboxParam}.
In this section we discuss some properties of the basis and of the
differential equation, as well as the numerical checks we have performed.

\subsection{Differential Equation and Alphabet}

The vector of master integrals contains 202 integrals, corresponding
to insertions on 126 different propagator structures.
We organise our basis so that the simpler topologies 
(the sunrise topologies, with 3 propagators) appear in the first entries, 
and the topology with most propagators (the hexa-box) appears in the
last entry.

Out of the 202 master integrals, 185 correspond to master integrals
that appear in five-point one-mass topologies. The remaining 17 integrals
are genuine 6-point massless integrals, and they appear as follows
(the indices given below correspond to the position
of the integrals as they appear in our ancillary files~\cite{zenodo}):
\begin{itemize}
  \item Penta-triangle (Fig.~\ref{fig:ptTopology} and permutation): a single master integral for each permutation, at positions \{149\} and \{150\}.
  The insertion in Eq.~\eqref{eq:pureCandidate_pt} gives a pure integral.
  \item Hexa-bubble (Fig.~\ref{fig:hbubTopology}): a single master integral at position \{154\}. 
  The insertion in Eq.~\eqref{eq:pureCandidate_hbub} gives a pure integral.
  \item Double-box (Fig.~\ref{fig:dbTopology}): seven master integrals at positions \{180,181,182,183,184,185,186\}.
  Insertions $\mathcal{N}^{\textrm{(db)}}_3$, $\mathcal{N}^{\textrm{(db)}}_4$, 
  $\mathcal{N}^{\textrm{(db)}}_5$ and $\mathcal{N}^{\textrm{(db)}}_6$ 
  in Eq.~\eqref{eq:pureCandidate_db} had to be corrected with the 
  procedure outlined in the previous section.
  The other insertions give pure integrals.
  \item Penta-box (Fig.~\ref{fig:pbTopology} and permutation): three master integrals for each permutation, at positions \{187,188,189\} and 
  \{190,191,192\}.
  Insertion $\mathcal{N}^{\textrm{(pb)}}_3$ in Eq.~\eqref{eq:pureCandidate_pb} had to 
  be corrected with the procedure outlined in the previous section. 
  The other insertions give pure integrals.
  \item Hexa-box (Fig.~\ref{fig:hbTopology}): a single master integral at position \{202\}.
  The insertion in Eq.~\eqref{eq:pureCandidate_hb} gives a pure integral.
\end{itemize}
Explicit expressions for all master integrals can be found in our ancillary
files~\cite{zenodo}. We stress once more that although the hexa-box decouples
for gauge-theory amplitudes in four dimensions, we included it in our basis
because it is particularly simple to find a pure integral from leading
singularity analysis, and doing so allowed us to explicitly check the
argument given in section~\ref{sec:9PropagatorIrrelevance}.

The other important quantity for the canonical differential equation is the
set of d$\log$ forms that appear, also known as the alphabet. 
We find that the dimension of the alphabet is 128, out of which only 11 
letters are associated with genuine six-point integrals. 
The new letters appear in the genuine six-point master integrals as follows:
\begin{itemize}
  \item Penta-triangle (Fig.~\ref{fig:ptTopology} and permutation): 
  The letters corresponding to each permutation appear at positions 
  \{118,123,125,127\} and \{119,123,126,128\}. The letter at 
  position \{127\} and its permutation, at position \{128\}, do not contribute 
  to the on-shell block of the differential equation.
  \item Hexa-bubble (Fig.~\ref{fig:hbubTopology}): The only genuine
  six-point letter appears at position \{123\}, and contributes to the on-shell
  differential equation.
  \item Double-box (Fig.~\ref{fig:dbTopology}): The only genuine
  six-point letters appear at positions \{120\} and \{123\}, 
  and they both contribute to the on-shell differential equation.
  \item Penta-box (Fig.~\ref{fig:pbTopology} and permutation): The letters corresponding to each permutation appear at positions 
  \{118,120,121,123,125,127\} and \{119,120,122,123,126,128\}. The letter at
  position \{121\} and its permutation, at position \{122\}, contribute to the 
  on-shell differential equation.
  \item Hexa-box (Fig.~\ref{fig:hbTopology}): The only genuine
  six-point letters appear at positions \{123\} and \{124\}, 
  and they both contribute to the on-shell differential equation..
\end{itemize}
In summary, we see that most of the genuine six-point letters can be
determined from the on-shell differential equations of the genuine six-point
topologies \cite{Henn:2021cyv}. 
The only exceptions are the pair of letters at position
\{127,128\}, which are related by permutation, and only appear in 
off-shell parts of the penta-triangle and penta-box differential equations.

The various letters of the alphabet can be classified with respect to their
parity with respect to the square-roots appearing in the definition of the 
pure basis. There are three square roots associated with three-point
Gram determinants,
\begin{align}\begin{split}\label{eq:threePointRoots}
  \lambda(p_{12}^2,p_{34}^2,p_{56}^2)=&\,
  s_{12}^2 + (s_{34} - s_{56})^2 - 2 s_{12} (s_{34} + s_{56})\,,\\
  \lambda(p_{16}^2,p_{23}^2,p_{45}^2)=&\,
  s_{16}^2 + (s_{23} - s_{45})^2 - 2 s_{16} (s_{23} + s_{45})\,,\\
  \lambda(p_{14}^2,p_{23}^2,p_{56}^2)=&\,
  s_{123}^2 + 2 s_{123} s_{234} + s_{234}^2 - 4 s_{23} s_{56}\,,
\end{split}\end{align}
see the definition of $\lambda(a,b,c)$ below Eq.~\eqref{eq:lambdaDef},
and the square root in Eq.~\eqref{eq:delta5} associated with the
five-point Gram determinant.  Out of the new genuine six-point letters
only the ones at positions \{127,128\} are odd under changing
the sign of the five-point Gram square root. As is already known, the
alphabet corresponding to five-point one-mass topologies includes
letters that are odd under the change of sign of the roots of
Eq.~\eqref{eq:threePointRoots} as well as the five-point Gram square
root.  There are also letters that are odd under the change of sign of
either the three-point and five-point Gram roots, and even under a
simultaneous change of both signs.  Explicit expressions for our
alphabet can be found in our ancillary files~\cite{zenodo}.

Finally, we note that we can close the alphabet under cyclic permutations
of the external legs, which gives the full alphabet for planar six-point
processes at two loops. We find that this alphabet closure contains 245
independent letters, and we also include them in our ancillary 
files~\cite{zenodo}.

\subsection{Numerical Checks}
\label{sec:numerics}

The differential equations we have discussed in the previous sections
can be used to obtain numerical evaluations of the corresponding master
integrals. In order to do so one needs to determine a boundary
condition. If the phase-space point corresponding to this boundary
condition is well chosen the integrals will evaluate to known numbers
and the boundary condition can be obtained analytically
(see e.g.~\cite{Henn:2024ngj} for such a point for six-point integrals).
In practice, however, in order to efficiently solve the integrals numerically
there is no particular advantage in having an analytic boundary 
condition \cite{Abreu:2023rco}.
Instead, high-precision numerical evaluations of the integrals can be
obtained with the \textsc{AMFlow} approach.  These can then be used as
numerical boundary condition to solve the differential equations with
generalised power series expansions~\cite{Moriello:2019yhu} or in
terms of Chen iterated integrals 
\cite{Chicherin:2020oor,Chicherin:2021dyp,Abreu:2023rco}.

In this section we describe how we
obtained numerical results from \textsc{AMFlow}, and how we compared
them to the numerical solution of our differential equations.  This
provides an important and completely independent check of the
correctness of our differential equations.
We leave the construction of a solution in terms of Chen iterated 
integrals that can be efficiently evaluated throughout phase-space 
for future work. Such a construction will be important for 
phenomenological applications of our results.

\paragraph{Boundary conditions with AMFLow:} We start by discussing the evaluation of our pure basis at two
benchmark points with \textsc{AMFlow}.
In order to construct rational four-dimensional points we start from a
twistor parametrisation of six-point kinematics (we use the twistor
variables $x_{i}$, $i=1,\dots,8$ given in~\cite{Henn:2021cyv},
and reported in appendix~\ref{app:twistor}).  We choose a point in
Euclidean kinematics, whose representation in Mandelstam variables
does not require too large rational numbers to avoid the proliferation
of large intermediate expressions in \textsc{AMFlow}.
Explicitly, we choose the point $P_1$ to be
\begin{align}\begin{split}
  P_1=&\,\big\{x_1 = -1, x_2 = -24, x_3 = 9, x_4 = 54, x_5 = 38, 
  x_6 = -97, x_7 = 3, x_8 = 95\big\}\\
  =&\,
  \big\{s_{12} = -1, s_{23} = -38, s_{34} = -\frac{1039}{6}, 
  s_{45} = -2712776, s_{56} = -50409, \\ 
  &\,\,\,s_{16} = -1662120, s_{123} = -95, s_{234} = -19926, 
  s_{345}  -2752175\big\}\,,
\end{split}\end{align}
and the point $P_2$,
\begin{align}\begin{split}
  P_2=&\,\big\{x_1 = -1, x_2 = -74, x_3 = 7, x_4 = 53, x_5 = 34, x_6 = -68, 
  x_7 = -91, x_8 = 76\big\}\\
  =&\,\big\{s_{12} = -1, s_{23} = -34, s_{34} = -\frac{827628}{37}, 
  s_{45} = -7995952, s_{56} = -147756, \\ 
  &\,\,\, s_{16} = -4804450, s_{123} = -76, s_{234} = -56882, 
  s_{345} = -6669452\big\}\,.
\end{split}\end{align}

To evaluate our pure basis with \textsc{AMFlow}, we collect
all integrals of the form of Eq.~\eqref{eq:topologyDef}
that appear in our pure basis before IBP reduction.
We then use \textsc{AMFlow} for the numerical evaluation of this set of
integrals, and during
the auxiliary-mass-flow procedure we explicitly instruct \textsc{Kira}
to express the IBPs in a basis obtained using the algorithms outlined
in~\cite{Smirnov:2020quc,Usovitsch:2020jrk}. This choice guarantees to
obtain IBP relations whose coefficients are rational, and have denominators
with a factorised dependence on $\epsilon$ and the Mandelstam
invariants.
This reduces the complexity of the rational coefficients and in turn
allows \textsc{Kira} to analytically reconstruct the
auxiliary-mass-flow differential equation with a small number of
finite fields of $64$-bit prime numbers.

We perform two simple consistency checks on the \textsc{AMFlow}
results. As a first check, we verify that the numerical calculation
for the pure basis starts at ${\cal O}(\epsilon^0)$, consistently with
the normalisation we chose for the master integrals.
Secondly, by imposing the so-called first-entry condition we can solve
the integrals up to order $\epsilon^1$ analytically and we verify that
the resulting solution agrees with that of \textsc{AMFlow}.  Both
numerical checks are passed at the two benchmark points $P_1$ and
$P_2$. We report the numerical evaluations in the ancillary
files~\cite{zenodo} with $20$ significant digits.

Before turning to the solution of our four-dimensional differential
equation, we close with a comment on the \textsc{AMFlow} evaluations.
In this procedure, we did not impose that the external kinematics
should be four dimensional, and the \textsc{AMFlow} calculation is
thus done with a full $D$-dimensional setup. The four-dimensionality
constraint only enters by requiring the numerical evaluations to be
performed at four-dimensional phase-space points. Since the
four-dimensional limit for the external kinematics is smooth (see
e.g.~\cite{Henn:2024ngj} for a discussion precisely in the context of
six-point kinematics), this procedure is well defined.

\paragraph{Numerical solution with \textsc{DiffExp}:}
The two numerical evaluations we obtained with \textsc{AMFlow} can be
used as a stringent independent check of the analytic differential
equation we constructed in previous sections. Indeed, we can take the
numerical result at $P_1$ as an initial condition, use the
differential equation to transport it to $P_2$, and verify that we
find agreement with the \textsc{AMFlow} result.

When using our differential equations, care must be taken to preserve the
four dimensionality constraint of the external kinematics, which is not explicit
in our expressions.
More precisely, if one takes a straight
line connecting the points $P_1$ and $P_2$ in Mandelstam space, it is not
clear that all points along this line will satisfy $\Delta_6=0$. However,
our differential equations are only valid on that surface.
A simple way to ensure that we do not leave the $\Delta_6=0$ surface is to
rewrite the differential equation in terms of twistor variables
(cf.~appendix~\ref{app:twistor}) and use the differential equation to evolve
between two points in twistor
space.\footnote{Alternatively, one can solve the $\Delta_6=0$
  condition for $s_{16}$ at the cost of introducing an explicit
  square root in the differential equation. This would introduce nested
  square roots in the letters of the differential equation, and so this
  route cannot be handled by \textsc{DiffExp}. Hence we proceed with the
  differential equation in twistor space.} This transformation is
trivial to perform on a differential equation written in terms of
$\mathrm{d}\log$-forms, and we can then solve it in terms of generalised power
series using the automated code \textsc{DiffExp}.

When comparing the \textsc{DiffExp} results to those of
\textsc{AMFlow}, one needs to account for a technical
subtlety. Specifically, we must be sure that we choose the branches of
the square roots in a consistent way. This is not guaranteed because
in one case the evolution is done in twistor space and in the other
the evolution is done in Mandelstam space.  In particular, since the
twistor variables rationalise $\sqrt{\Delta_5}$ this quantity might
change sign along the evolution path and become inconsistent with the
branch chosen in Mandelstam space. Pure integrals that are normalised
by such roots will be sensitive to this branch choice through an
overall sign change, so we must keep track of such signs when
comparing the two numerical evaluations.

Starting from $P_1$ as a boundary condition, we find perfect agreement
between the \textsc{DiffExp} and \textsc{AMFlow} results at
$P_2$. This is a highly non-trivial check, given that while
\textsc{DiffExp} is performing the evolution directly on the
four-dimensional hyper-surface, the evaluation of \textsc{AMFlow} at
the four-dimensional point $P_2$ is obtained via the
auxiliary-mass-flow procedure in a full $D$-dimensional setup. 
As already argued above,
the agreement between the two solutions is compatible with the fact that
the limit where the external kinematics are taken to be in four dimensions
is smooth~\cite{Henn:2024ngj}.

%%%%%%%%%%%%%%%%%%%%%%%%%%%%%%%%%%%%%%%%%%%%%%%%%%%%%%%%%%%%%%%%%%%%%%%%%%%

%%%%%%%%%%%%%%%%%%%%%%%%%%%%%%%%%%%%%%%%%%%%%%%%%%%%%%%%%%%%%%%%%%%%%%%%%%%

%%%%%%%%%%%%%%%%%%%%%%%%%%%%%%%%%%%%%%%%%%%%%%%%%%%%%%%%%%%%%%%%%%%%%%%%%%%

\section{Conclusions}

In this work, we have constructed and solved the differential equations for the
collection of planar two-loop massless six-point master integrals
entering scattering amplitudes in four-dimensional gauge theories.
As the external momenta of the problem are linearly dependent, this introduces
novel challenges for the differential equation strategy. We choose to tackle
them head on, 
avoiding the complications introduced by four-dimensional
parametrizations such as those constructed from momentum twistors. 
Instead, we work directly in Mandelstam
space where the invariants satisfy a high-degree polynomial constraint given by
the vanishing of the 6-point Gram determinant.

Before going into the details of the calculation of the differential equations
for six-point two-loop planar integrals, an important result of this
paper is that we show that we do not actually need to compute all the 
six-point topologies
if we are interested in applications to four-dimensional gauge theories,
such as e.g.~the calculation of four-jet production at hadron
colliders.
We reduced the number of six-point topologies that we need to investigate
by arguing that the six-point topologies with 9 propagators
do not need to be computed. 
Indeed, we showed that they can be related to evanescent
integrals plus a linear combination of integrals with a lower number of
propagators, analogous to the pentagon at one loop.

To tackle the calculation of the topologies that do need to be computed,
we started by discussing how we construct differential operators which
naturally take into account the four-dimensionality constraint of the
external kinematics.
Next, we introduced a novel analytic reconstruction technique for
the canonicalization of the differential equation.
Specifically, we use numerical evaluations of the $\epsilon^0$ part of the
pre-canonical differential equation to directly reconstruct the change of basis
matrix, by exploiting the expectation that the change of basis matrix is
algebraic. This allows to both dramatically decrease the number of samples 
required, as well as bypass the integration step. 
This setup is generic, and we believe applicable to other multi-scale problems.
In our application, it had to be implemented taking
into account the constraint following from the vanishing of the 6-point 
Gram determinant, and we explained how we handled the subtleties associated
with this extra complication.
Once a pure basis had been obtained, determining the associated differential
equations proved to be trivial using techniques that are now standard.
Finally, we compute a benchmark evaluations using \textsc{AMFlow}, 
and discussed how to solve the differential 
equation using public tools such as \textsc{DiffExp}.
The fact that these two approaches agree provides a non-trivial check of
our results.

We expect that the analytic reconstruction technology that we have introduced
for obtaining a canonical basis will be readily applicable to other six-point
integrals, and to be a valuable tool for problems of higher multiplicity
or with more scales. 
Beyond such applications, we also expect that the observed decoupling of the
9-propagator topologies in the $D \rightarrow 4$ limit is a specific example of
a much more general phenomenon.
It would be of great interest to understand if the decoupling could be argued
for in full generality.

While we leave the implementation of an efficient code for the
numerical evaluation of these integrals for future work, this study
of the complete set of two-loop six-point planar integrals relevant for
four-dimensional gauge theories provides a crucial step towards the calculation
of important scattering processes, such as the production of four jets at
hadron colliders at NNLO accuracy.

\begin{acknowledgments}
  P.F.M is funded by the European Union (ERC, grant agreement
  No. 101044599).
  B.P. received funding from the European’s Union Horizon 2020
  research and innovation programme LoopAnsatz (grant agreement number
  896690) and by the European Union (ERC, MultiScaleAmp, Grant
  Agreement No. 101078449).
  J.U. is funded by the Deutsche Forschungsgemeinschaft (DFG, German
  Research Foundation) Projektnummer 417533893/GRK2575 “Rethinking
  Quantum Field Theory”and by the European Union through the European
  Research Council under grant ERC Advanced Grant 101097219 (GraWFTy).
  Views and opinions expressed are however those of the authors only
  and do not necessarily reflect those of the European Union or the
  European Research Council Executive Agency. Neither the European
  Union nor the granting authority can be held responsible for them.
\end{acknowledgments}

\newpage
\appendix

\section{Twistor variables representation of Mandelstam invariants}
\label{app:twistor}

In this appendix we provide a parametrisation of the Mandelstam
invariants featuring in the six-point problem in terms of twistor
variables. We use this representation in the numerical solution of the
differential equation in four-dimensional external kinematics using
the \textsc{DiffExp} code (see Sec.~\ref{sec:numerics}).
The following parametrisation is taken from Ref.~\cite{Henn:2021cyv}
\begin{align}\begin{split}\label{eq:twistor}
  s_{12} &= x_1\,,\\
  s_{23} &= x_1 x_5\,,\\ 
 s_{34} &= -((
   x_1 (x_2 x_3 x_6 + x_5 (-1 + x_3 (-1 + x_2 (-1 + x_7)))))/
          x_2)\,,\\
  s_{45} &= 
  x_1 (-((1 + x_4 + x_3 x_4) x_5 (-1 + x_7)) \\&\quad- (1 + x_3) x_4 x_8 + (x_2 x_3 x_4 (x_5 (-1 + x_7) + x_6 x_8))/
     x_5)\,,\\ s_{56} &= (
  x_1 x_3 (-x_5 (x_5 (-1 + x_7) + x_8) + 
     x_2 (x_5 (-1 + x_7) + x_6 x_8)))/x_5\,,\\ 
 s_{16} &= -((x_1 x_2 x_3 x_4 (x_5 (x_6 - x_7) - x_6 x_8))/
          x_5)\,,\\
  s_{123} &= x_1 x_8\,,\\ 
 s_{234} &= x_1 x_3 (x_2 x_6 - x_5 x_7)\,,\\ 
 s_{345} &= x_1 (x_6 + 
     x_4 (-1 + x_6 + 
        x_3 (-1 + x_6 + x_2 (-1 + x_7 + (x_6 x_8)/x_5))))\,.
\end{split}\end{align}

\bibliographystyle{JHEP}
\bibliography{main}

\end{document}